\documentclass[final,3p,times]{elsarticle}

\usepackage{longtable}
\usepackage{amsmath,amsfonts}
\usepackage{algorithmic}
\usepackage{algorithm}
\usepackage{array}
\usepackage{url}
\usepackage{textcomp}
\usepackage{stfloats}

\usepackage{verbatim}
\usepackage{booktabs} 
\usepackage{graphicx}
\usepackage{subcaption}
\usepackage{tabularx}
\usepackage{booktabs}
\usepackage{hyperref}

\begin{document}
	
	\begin{frontmatter}
		
		\title{Big Data Energy Systems: A Survey of Practices and Associated Challenges}

		\author [unile] {Lunodzo J. Mwinuka}
		\ead{lunodzo.mwinuka@unisalento.it}
		\author [unile] {Massimo~Cafaro\corref{cor1}}
		\ead{massimo.cafaro@unisalento.it}
		\cortext[cor1]{Corresponding author}
		\author [tecnico] {Lucas Pereira}
		\ead{lucas.pereira@tecnico.ulisboa.pt}
		\author [tecnico] {Hugo Morais}
		\ead{hugo.morais@tecnico.ulisboa.pt}
		
		\affiliation[unile]{organization={University of Salento, Dept. of Engineering for Innovation},
             addressline={Via per Monteroni},
             city={Lecce},
             postcode={73100},
             country={Italy}}
             
             \affiliation[tecnico]{organization={Instituto Superior Técnico, Universidade de Lisboa},
             addressline={av. Rovisco Pais 1},
             city={Lisbon},
             postcode={1049-001},
             country={Portugal}}

		\begin{abstract}
			Energy systems generate vast amounts of data in extremely short time intervals, creating challenges for efficient data management. Traditional data management methods often struggle with scalability and accessibility, limiting their usefulness. More advanced solutions, such as NoSQL databases and cloud-based platforms, have been adopted to address these issues. Still, even these advanced solutions can encounter bottlenecks, which can impact the efficiency of data storage, retrieval, and analysis. This review paper explores the research trends in big data management for energy systems, highlighting the practices, opportunities and challenges. Also, the data regulatory demands are highlighted using chosen reference architectures. The review, in particular, explores the limitations of current storage and data integration solutions and examines how new technologies are applied to the energy sector. Novel insights into emerging technologies, including data spaces, various data management architectures, peer-to-peer data management, and blockchains, are provided, along with practical recommendations for achieving enhanced data sharing and regulatory compliance. 
		\end{abstract}
		
%


	\end{frontmatter}
	
	\section{Introduction}
	The current nature of data is intricate, characterised by a high growth rate in terms of volume, among other key attributes. This is witnessed by the rapid growth of generated data (e.g. in 2023 and 2024, a total of 120 and 147 zettabytes (ZBs) were generated, respectively). In addition, it is estimated that there will be a 23.13\% increase in 2025 \cite{RN102}. In energy systems, one reason for this is the huge number of connected devices that generate data in real-time \cite{RN45}, and this number is growing \cite{RN98}. It is also observed that the deployment of data acquisition technologies is advancing rapidly \cite{RN276}, resulting in an exponential growth of energy data. 
	
\begin{figure*}
		\centering
		\includegraphics[width=.8\linewidth]{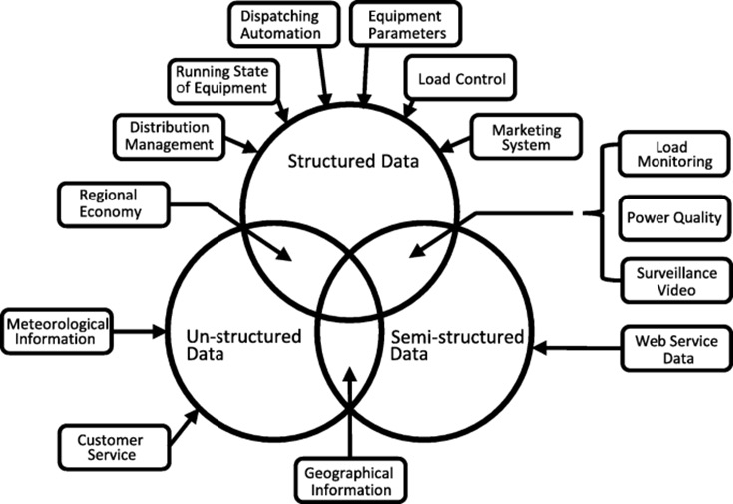} 
		\caption{Data sources in smart grids \cite{RN65}.}
		\label{fig:data_sources}
	\end{figure*}

	Energy systems involve the integration of billions of digital devices that communicate and exchange data globally \cite{RN124}. These devices collect data in real-time to support the operation and management of the systems. Devices like smart meters collect the status of the electricity grid to aid in identifying supply interruptions, inefficient voltages, incorrect connections, and energy supply and consumption, thereby maintaining a balance between supply and demand. Sensors, on the other hand, monitor the system's operation and conditions to minimise downtime risks, among other benefits. In this case, more than 1 billion smart meters were already deployed in 2022, and approximately 13 billion connected devices with automated controls and sensors were actively used in 2023, thereby generating a substantial amount of data \cite{RN45, SmartM}. Smart grids stand at the core of digital advancements in the energy sector, generating and managing a wide variety of data. The sources of data in the smart grids can be grouped into three main categories: \textit{structured}, \textit{semi-structured} and \textit{unstructured} data, with each category being strictly related to the stratified nature of data depending on both the systems' operation and machines involved in the production, transmission and distribution of energy \cite{RN65}. An overview of the data sources in smart grids is shown in Figure \ref{fig:data_sources}.
	
	Unfortunately, the majority of the generated data remains unused, thereby losing its potential. For instance, only 2\% to 4\% of smart meter-generated data is being used to enhance the efficiency of grid operation \cite{RN45, sevilla2022state}. Furthermore, the efficient utilisation of generated data is not realised due to challenges in the interoperability of multiple data management layers, as well as difficulties in data extraction. The global transition to renewable energy and the advancement in smart grid technology usage demand innovative data management approaches. With a growing number of connected devices that generate diverse datasets, addressing the challenges of data integration, accessibility, and scalability is more critical than ever.
	
	\subsection{Motivation}
	Uncovering the value of data in energy systems is crucial for facilitating smooth system operations, among other benefits. This review examines current big data management practices in energy systems and other sectors between 2016 and 2025 and channels the findings toward future directions. Despite current research being short of reviews that look at energy systems and big data management, previous surveys have focused on general big data deployment to explore the state-of-the-art practices \cite{RN123, RN58,freitas2025data, G2025102408}, big data storage \cite{RN104, RN128, RN134, RN132}, data mining \cite{RN137}, and distributed database technologies for big data \cite{RN198}. State-of-the-art reviews focusing on big data technologies, examining their trends and current limitations in energy systems are scarce. Some, like \cite{RN123, RN58, jaramillo2025bibliometric}, have left room for exploration in big data management approaches. Others \cite{RN104, RN128, RN134, RN132} have focused on health, transport, industries, and other sectors \cite{RN123, RN112, RN113, RN118, RN122, RN110, RN111, RN116, hajjaji2021big, RN115}. 
	
	One of the recurring issues in the existing reviews is the lack of exploration of data integration strategies. While some studies have provided a general overview of big data technologies and their application in smart grids \cite{RN123, RN104, RN113, RN118, hajjaji2021big, RN258, RN413, RN114, RN424, RN195}, they often fall short of addressing the complexities of data integration, particularly when dealing with distributed storage systems. This gap leaves significant challenges in harmonising diverse data sources, which is crucial for effective decision-making in energy systems management and improving data sharing.
	
	Another key limitation is the focus on specific aspects of big data technologies, such as data analytics solutions or data mining, without providing a holistic view of the complete data management lifecycle. For instance, many reviews emphasise the evolution of storage technologies from traditional relational databases to NoSQL solutions. Yet, they do not fully explore how these technologies can be integrated into the broader energy sector. Challenges such as frequent data updates, data partitioning, and replication, which are already prevalent in the energy domain, remain unexplored. This highlights the need for a more robust review that can simultaneously give an overview of the possibilities for achieving data quality, consistency, availability, and integration. 
	
	Some reviews also prioritise cloud-based storage strategies, which, while addressing scalability concerns, introduce new challenges without recommendations for possible alternatives. Moreover, some studies fail to account for the unique requirements of energy systems, such as regulatory compliance and specific architectural needs for deploying big data solutions. This lack of sector-specific insights often results in recommendations that are not fully applicable or effective for energy data management.
	
	Additionally, while advanced analytical tools like Hadoop are frequently discussed, there is limited focus on how these tools can be effectively utilised for data management and storage within energy systems. The emphasis often leans towards analytics, leaving critical aspects such as data storage, retrieval, and integration underexplored. This gap suggests a need for more comprehensive exploration that not only focuses on analytics but also on the foundational elements of the big data management lifecycle, i.e., data storage and integration. A summary of identified gaps for exploration and current reviews is presented in Table \ref{tab:gap}.

	\begin{table*}[t]
		\centering
		\begin{tabular}{|c|c|c|c|c|c|c|c|}
			\hline
			\textbf{Reference} & \textbf{Year} & \multicolumn{6}{c|}{\textbf{Explored domains or Addressed challenges}} \\
			\cline{3-8}
			&  & \textbf{Energy} & \textbf{NoSQL} & \textbf{Storage} & \textbf{ Decentralised} & \textbf{Cloud} & \textbf{Data Integration} \\
			\hline
			{\cite{RN58}} & 2016 & Yes & Yes & Yes & No & No & No \\
			\hline
			{\cite{RN134}} & 2017 & No & Yes & Yes & No & No & No \\
			\hline
			{\cite{RN128}} & 2019 & No & Yes & Yes & Yes & Yes & No \\
			\hline
			{\cite{RN114}} & 2021 & Yes & No & No & No & No & No \\
			\hline
			{\cite{RN104}} & 2021 & No & Yes & Yes & No & No & Yes \\
			\hline
			{\cite{RN137}} & 2021 & Yes & No & No & Yes & No & No \\
			\hline
			{\cite{RN123}} & 2023 & No & No & No & No & No & No \\
			\hline
			{\cite{RN135}} & 2023 & Yes & No & No & No & No & No \\
			\hline
			{\cite{RN97}} & 2023 & Yes & No & No & No & No & No \\
			\hline
			{\cite{RN258}} & 2024 & No & No & No & No & No & No \\
			\hline
			\cite{freitas2025data} & 2025 & No & No & Yes & No & Yes & No \\
			\hline
		\end{tabular}
		
		\caption{An overview of gaps and limitations in the current literature reviews.}
		\label{tab:gap}
	\end{table*}

	This work discusses the limitations of current storage and integration solutions, as well as the application of new technologies within the energy sector. Additionally, practical recommendations are provided to enhance the efficiency of these technologies in energy systems and related areas, to improve data sharing, quality, accessibility, scalability, and overall performance of storage systems. To this end, this work can be considered as a guidebook for novel approaches to the deployment of big data technologies based on two major big data areas: (i) storage and (ii) integration. Hence, our contribution can be summarised as follows:
	
	\begin{itemize}
		\item We identify key trends in big data, particularly the data management needs and applications in energy systems, and highlight the regulatory frameworks guiding the management of data;
		\item We explore the current methods for managing (with a focus on storage) and integrating large-scale data, discussing existing practices and the challenges they face;
		\item We outline the critical challenges in big data management for energy systems and propose potential solutions, drawing on successful practices from other fields.
	\end{itemize}
	
	\subsection{Methods}
	\label{subsec:methods}
	Given the context of this review, the exploration leveraged popular academic search engines, databases and websites that publish trends in data management and energy systems. Diverging from the typical systematic literature review process, this review adapts patterns of rapid literature review methods and narrative literature review methods proposed in \cite{RN99, RN140, RN153}. This is done to gather evidence from the current practices. Specifically, this review adopts a quick scoping approach to respond to this work’s questions by mainly consulting the Web of Science to query and filter sources, Google Scholar, IEEE Explorer, and journals’ websites for extracting actual files. This method was chosen for its efficiency in synthesising a large body of literature within a limited time frame, making it particularly suited for identifying emerging trends and current gaps.
	
	Relevant keywords that reflect the anticipated outcomes of this review were selected. Without a specific order, the study used “big data,” “big data management,” “big data management in energy systems,” “big data storage,” “big data storage in energy systems,” “big data integration,” “large-scale data management” and “large-scale data integration”. To better understand the state of the art, the review focused on more recent publications. Articles with theoretical and foundational knowledge of the underlying architectures were not subject to a strict time constraint, allowing for a comprehensive understanding of both theoretical and technical frameworks. 
	
\begin{figure*}
		\centering
		\includegraphics[width=.7\linewidth]{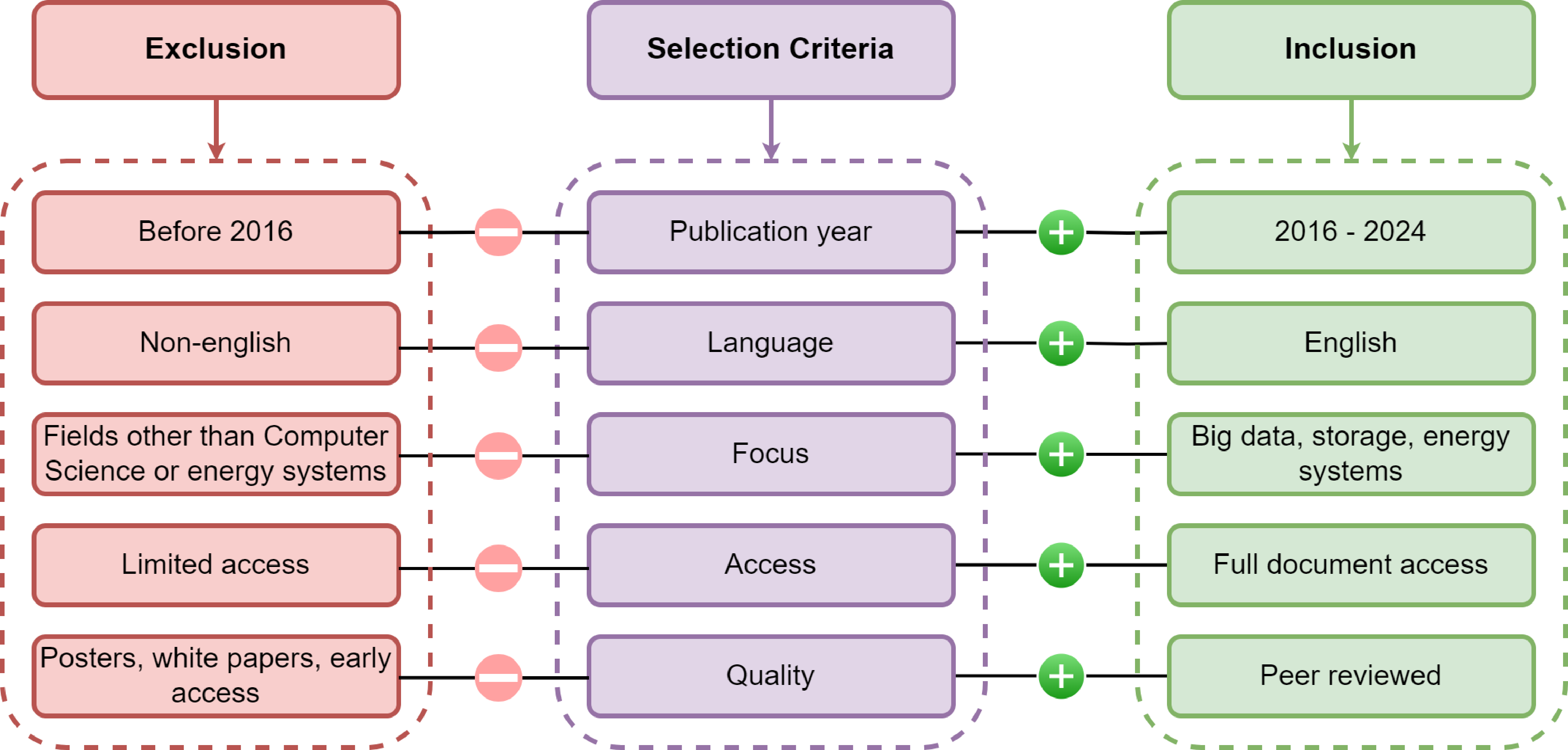}
		\caption{Study selection criteria.}
		\label{fig:study_selection}
	\end{figure*}

	After searching for the identified keywords, a total of 12,290 publications were found. For a more focused research work, inclusion and exclusion criteria were added, as summarised in Figure \ref{fig:study_selection}. This process resulted in a total of 298 research works, from which the titles and abstracts were screened to select the articles that best addressed the concerns of this review. Hence, the review was conducted using 53 articles, whose selection was guided by a comparison of titles and abstracts in relation to the identified review concerns.
	
	The remainder of this review is organised into four sections. Section \ref{sec:background} provides background information on big data to explain important terms that will be used throughout this work. It also sets the ground for understanding big data in the context of the energy domain. Section \ref{sec:data_mgt_state} provides an overview of the state-of-the-art response to existing data management approaches, regulatory guidelines, data storage and data integration practices. We propose our recommendations in Section \ref{sec:recommendation}, and draw our conclusions in Section \ref{sec:conclusion}.
	
	\section{Background Information}
	\label{sec:background}
	
	\subsection{Big Data Definitions}
	\label{subsec:bigdata_def}
	Data are generated in various formats and structures, hence the rise of the term "big data" without a formal definition \cite{RN45, G2025102408}. Several scholars have different perspectives on the definition of big data. Better yet, scholars agree that any data that comes in huge quantities has various structures, has the potential to generate knowledge, and cannot be managed by conventional databases, which can be termed big data. It is also commonly referred to as large-scale data. However, in large-scale data, emphasis is placed on the size or quantity of data. In this work, the terms are used interchangeably.
	
\begin{figure*}
		\centering
		\includegraphics[width=.7\linewidth]{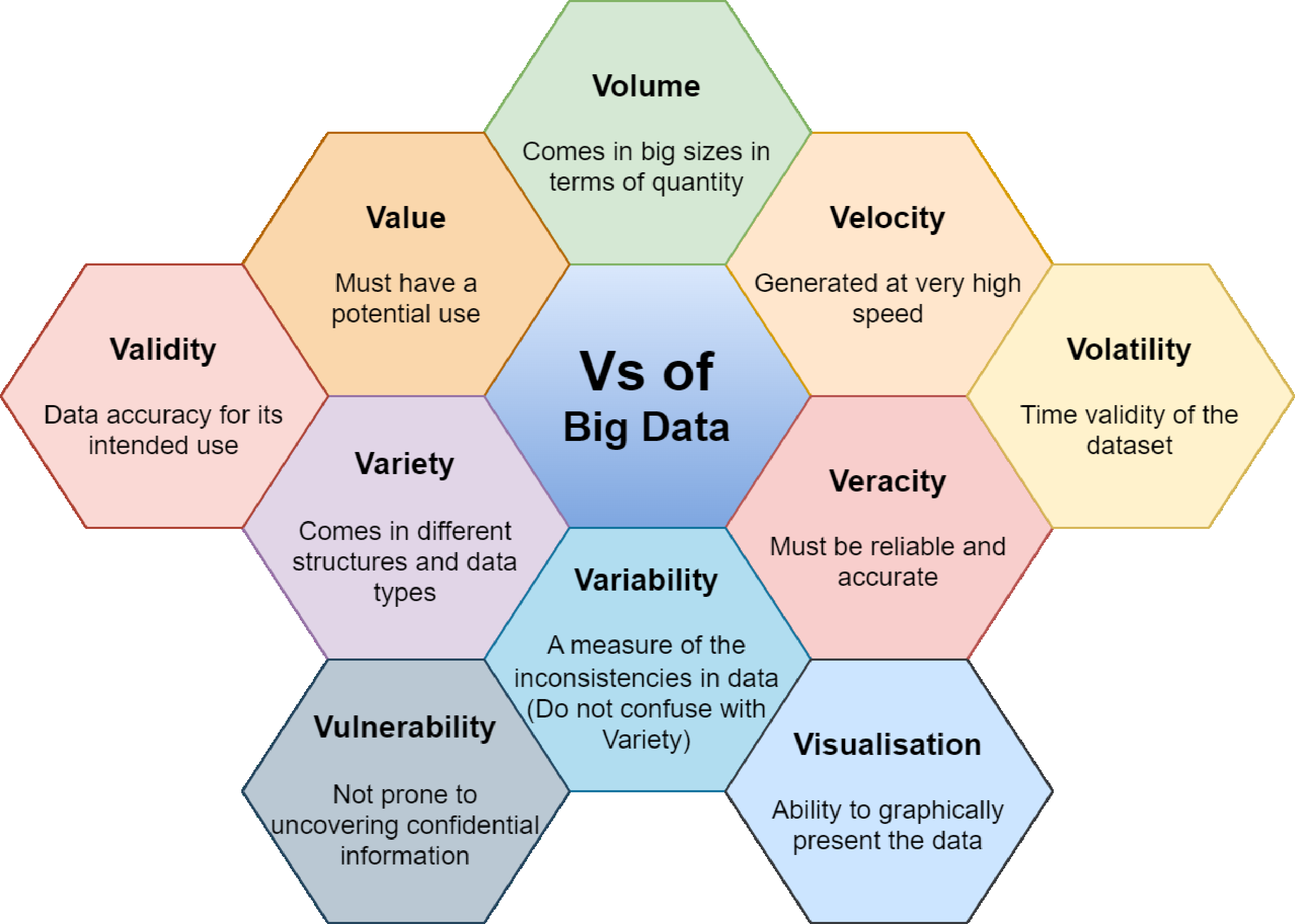}
		\caption{A presentation of 10 Vs of Big Data.}
		\label{fig:vs}
	\end{figure*}

	Commonly, big data is viewed as data characterised by several popular Vs. The initial definition considered big data to have three main features: Volume (size of data), Velocity (speed of generation and processing), and Variety (types of data), generally referred to as 3V’s \cite{RN56, RN46}. However, some scholars have sought to enrich it (4 Vs) by adding value to the definition and emphasising that data must have the potential for use and hold information that can be extracted \cite{RN100}. Authors in \cite{RN114} present the 5 Vs structure, adding Veracity, which emphasises data accuracy and reliability. The largest extension of Vs details 10 Vs, covering the 5 Vs and bringing Variability, Validity, Volatility, Vulnerability, and Visualisation as important aspects to consider in defining big data \cite{RN101}. This makes the definition of big data a relative concept rather than an absolute definition \cite{RN65}. A descriptive summary of the most common Vs is presented in Figure \ref{fig:vs}. 
	
	\subsection{Big Data Management}
	\label{subsec:bigdata_mgt}
\begin{figure*}[!t]
		\centering
		\includegraphics[width=.8\linewidth]{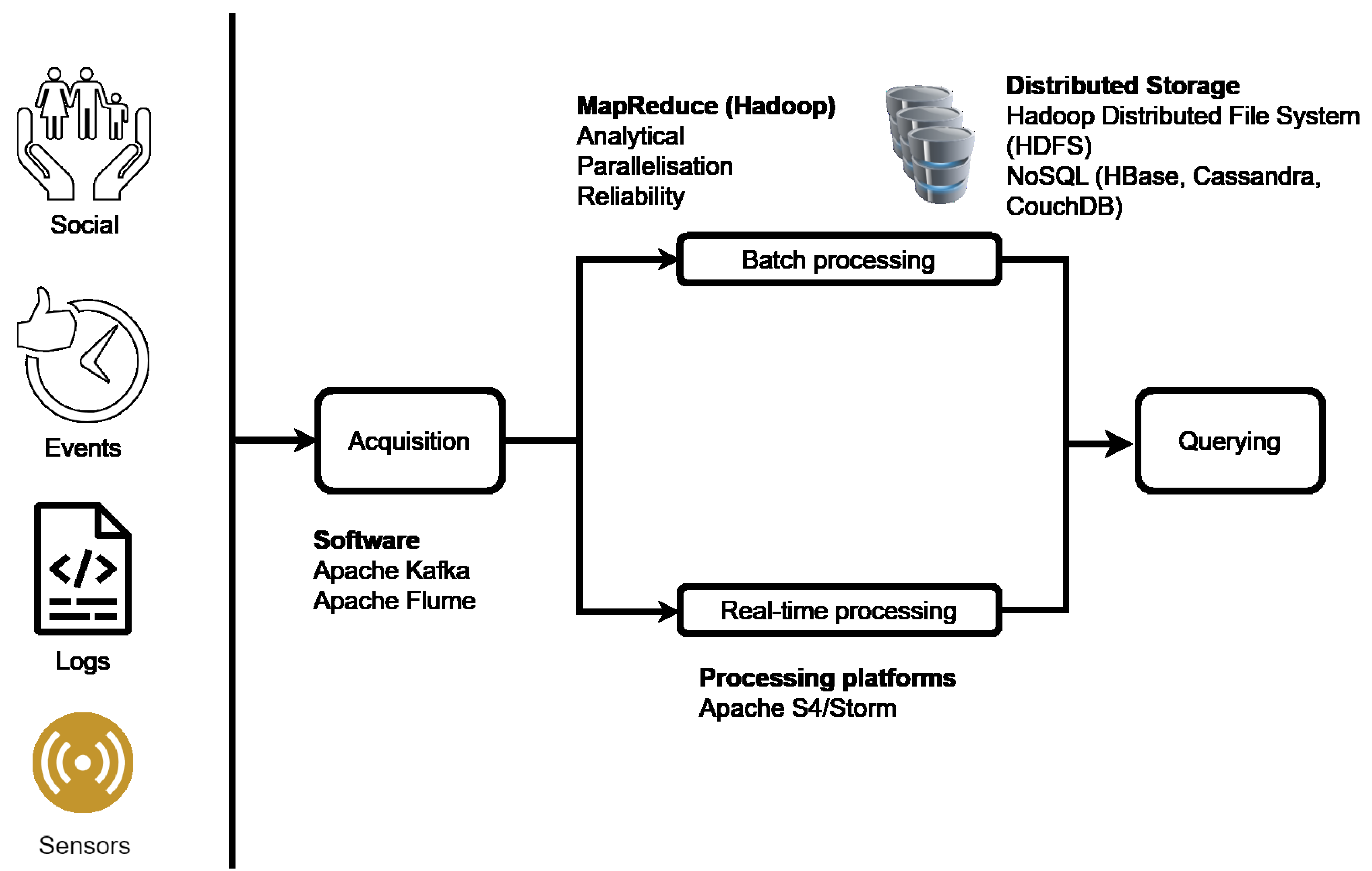}
		\caption{Big data workflow (customised from \cite{RN56}).}
		\label{fig:workflow}
	\end{figure*}

	Big data management entails all activities involved in the data lifecycle. Generally, activities in managing a big data stack begin with the acquisition of data. Data collected usually come from various sources, including social media, sensors, logs, and system events. Popular acquisition software options include Apache Kafka and Flume, depending on the types of data collected and the underlying computing infrastructure. The data acquisition is followed by data processing, which can be done in batch or real-time. Data processing is facilitated by tools such as MapReduce and Apache Storm, while Hadoop, through the Hadoop Distributed File System (HDFS), facilitates its storage. Lastly, data can be retrieved for different use cases through querying \cite{RN56}. The basic big data management workflow, as described, is depicted in Figure \ref{fig:workflow}. 
	
\begin{figure*}
		\centering
		\includegraphics[width=.8\linewidth]{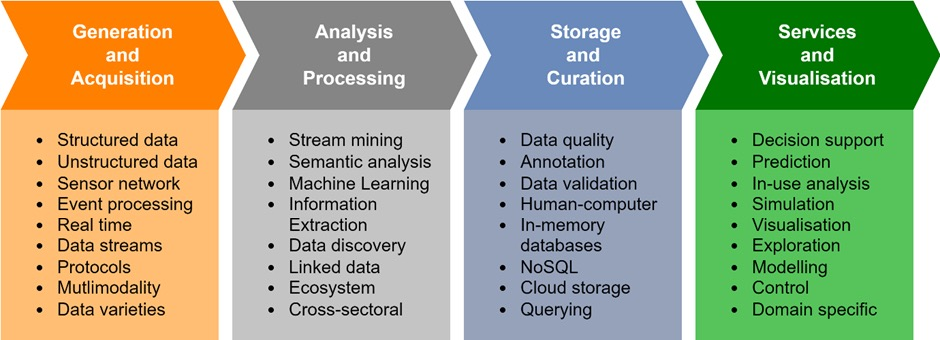}
		\caption{Big data value chain (customised from \cite{RN56}).}
		\label{fig:value_chain}
	\end{figure*}

	Alternatively, the flow of big data management can be looked at in terms of its value chain, which is our preferred approach. The big data value chain provides a structured approach to understanding and benefiting from data, thereby serving as an important component in the management and optimisation of big data (see Figure \ref{fig:value_chain}). The big data value chain can be grouped into four major categories: -
	
	\begin{itemize}
		\item First, the data acquisition layer: this layer collects both structured and unstructured data from varied sources. To efficiently collect data, several approaches are employed to support multimodality, including the real-time acquisition of data from multiple streams, such as sensor networks, IoT devices, logs, and smart meters. The collected data become an input for the data processing and analysis;
		
		\item Second, data analysis and processing: Data analysis plays a crucial role within the big data value chain, as it transforms raw data into actionable knowledge. This process benefits from several methods, such as machine learning (ML) \cite{RN194}, stream mining \cite{RN195}, semantic analysis \cite{RN196}, information extraction, and data discovery. Stream mining, for instance, processes continuous data streams in real-time, while ML algorithms apply semantic analysis to interpret the meaning and context of the data. Furthermore, the integration of datasets through linked data broadens the scope of analysis, enabling a comprehensive understanding of cyber-physical systems;
		
		\item Third, storage and curation: data curation is done through validation, provenance checks, computation and other methods to ensure the quality of data in storage systems. At this stage, an emphasis is put on maintaining data quality and trustworthiness, which is critical for the validity of any subsequent analysis. Then, the curated data is stored to make sure that it is accessible and ready for other uses, including knowledge extraction. At this stage, the longevity and accessibility of data must be the norm. Storage systems are commonly implemented using in-memory databases, NoSQL, cloud and other advanced approaches. Typically, cutting-edge technologies enable storage systems to provide robust and rapid access to large data sets \cite{RN197}. Cloud storage, in particular, provides scalability and sophisticated query interfaces that simplify data retrieval tasks \cite{RN221}. Standardisation and consistency are the watchwords here, guaranteeing that data remains coherent and interoperable across various applications and platforms. Key quality checks in storage systems focus on scalability, performance, consistency, availability, accessibility, sharing, security, and privacy, ensuring that data remains not only accessible but also reliable and secure for long-term use \cite{RN72}. The choices between these approaches may vary due to the specific requirements of the businesses;
		
		\item Fourth, Services and visualisation: in many cases, data are consumed for decision-making, prediction, analytics, simulation, visualisation, modelling and domain-specific usage, among many other use cases. This falls into the data usage chain of value. There are several use cases that businesses benefit from in this stage, including the drive to decision-making and the ability to make accurate predictions. Furthermore, through visualisation and modelling, complex datasets are translated into comprehensible formats, aiding stakeholders in making informed decisions.
	\end{itemize}
	
	With the current data trends, it is vital to innovate approaches to data (evidently, large-scale) management. An efficient management approach would consider the entire data ecosystem, with a focus on its value chain. A successful data ecosystem would bring together the stakeholders whilst designing the data management platform \cite{RN56}. Interaction between these stakeholders is vital to realising an efficient data-driven economy. For instance, authors in \cite{RN56} present a mapping of the requirements guiding a technological roadmap for fulfilling the key requirements along the data value chain for the energy and transport sectors. This mapping focused on technology that was not readily available but required further research and development.
	
	Major technical requirements focus on data sharing and access, real-time analytics, prescriptive analytics and platforms that facilitate abstraction. Whilst most technologies for analytics are getting concrete, the aspect of data sharing and access to facilitate data linkage and scalability is still a valid research concern. This is due to the need to combine data, including usage data and information on products and services, to enhance efficiency in sales and operations.
	
	\subsection{Data Availability in Energy Systems}
	\label{subsec:data_in_energy}
	Like in any other field, data is essential in energy systems. They span their potential from power generation to power supply and consumption. Its potential, among many others, includes aiding in the proper maintenance of systems, planning and error detection in machine and systems operation \cite{RN124}. Modern data in energy systems are collected from consumers' smart meters, Phasor Measurement Units (PMUs), Supervisory Control and Data Acquisition (SCADA) systems, and several new sensors installed in different devices and assets \cite{RN125}. In the context of power generation and distribution, these data can be used to aid in system planning, operation and maintenance by facilitating fault detection and ensuring smooth operations \cite{RN45, RN49}. This practice enhances efficiency in energy systems by collecting real-time data, which inform the current state of energy systems \cite{RN137}. However, with a growing number of data sources and their diverse nature, effective data acquisition, storage, and integration have become challenging whilst simultaneously being extremely crucial.
	
	Data from meter readings are usually collected periodically, depending on the demands for data and business policies. For instance, the Électricité de France (EDF) collects meter reading data once a month \cite{RN56}. Smart meters are one of the data collection points in energy systems. For electricity smart meters, data is usually collected every 15, 30, or 60 minutes, with a single reading accounting for a few kilobytes (KB), resulting in an average total of 100–200 KB. At the end of 2023, there were more than 186 million smart meters in Europe, an increase of 4\% from 2022. The penetration of smart meters in Europe is expected to increase from 60\% in 2023 to 78\% in 2028 \cite{RN139}. Smart meters collect data about electricity consumption in the location where they are installed, and with their increasing number, the data will also increase.
	
	Power grids are also generating enormous amounts of data, with around 320 million sensors deployed globally. The deployed sensors transmit real-time data from the grids. The collected data are not effectively utilised, with only a fraction of the data currently being used to enhance the efficiency of grid operation \cite{RN45}. Due to these trends, energy data exhibit all the characteristics of big data, necessitating advanced approaches for its acquisition, storage, integration, and processing.

	\section{Data Management: the state of the art}
	\label{sec:data_mgt_state}
	\subsection{An Overview}
	\label{subsec:overview}
	Most existing research on big data is tailored towards its challenges and potential for analytics, influenced by current data trends \cite{RN135, RN214, RN234, RN235, RN212, RN215, RN225, RN227, RN230, RN238, RN201, RN233, RN236, RN240}. The need to understand what is contained in the data is readily apparent in some research works \cite{RN118, hariri2019uncertainty, RN107}. Similarly, the need to address challenges related to data quality, sharing, storage, and integration is evident in most analytics research works, which demand advanced data cleaning techniques to improve their performance \cite{RN135, RN277}. The data preprocessing activities usually involve reducing or eliminating noise and inconsistencies, which are, in most cases, common in unstructured and semi-structured data. These activities become way more complicated if data is not handled well since the acquisition stage.
	
	Robust analytics frameworks are common and, of course, essential among researchers to realise the full potential of big data. These frameworks are designed to handle the data complexity, scale of data, and its varieties. Their strength relies on the ability to automate most tasks, including data cleaning, transformation, and, in some cases, data integration. This allows analysts to focus more on deriving insights than data preparations. Frameworks such as Apache Hadoop, Spark, and Kafka are often employed to process large-scale datasets efficiently \cite{maatallah2025real, RN237}. Their ability to manage parallel computations enables organisations to extract valuable insights more quickly and effectively, thereby enhancing decision-making in areas such as energy distribution, consumption optimisation, and predictive maintenance \cite{RN234, RN213}.
	
	One notable challenge is the uncertainty of the data. Uncertainty can stem from several factors, including sensor malfunctions, missing data, data disparities among data owners (which are often not integrated), and the existence of multiple sources \cite{RN278}. Among the practical ways to deal with these challenges is the use of probabilistic models and stochastic algorithms to accommodate variability and imprecision \cite{RN279}. Frameworks that incorporate uncertainty modelling help ensure that analytics remain reliable even when the data is incomplete or noisy. 
	
	In energy systems, this is particularly important for applications such as load forecasting and grid optimisation, to mention a few, where uncertain or inaccurate data can lead to suboptimal decisions. Techniques such as Monte Carlo simulations and Bayesian networks \cite{gui2011bayesian, salboukh2025reliability} can be used to model uncertainty and provide decision-makers with a range of potential outcomes, thereby improving the resilience of energy systems in dealing with fluctuating and unpredictable conditions \cite{RN253}.
	
	The literature also evidences that the value derived from big data in energy systems is not just limited to operational efficiencies; rather, it extends to strategic decision-making and long-term planning \cite{RN234, RN238}. Big data allows energy providers to gain a deeper understanding of consumption patterns \cite{RN234}, predict equipment failures, optimise supply chains \cite{RN113}, and even forecast the impact of integrating renewable energy sources into the grids. 
	
	Exploiting big data requires a shift from simply storing vast amounts of data to actively mining that data for actionable insights. However, these requirements must be facilitated by effective data acquisition and storage methods, which will then benefit the subsequent stages of the big data life cycle. Here, we further explore management approaches based on the data currently available in energy systems. Also, excerpts from the literature on data storage approaches and integration methods will be presented and discussed.
	
	\subsection{Data Management Approaches}
	\label{subsec:datamgt_approaches}
	Data management is important owing to its role in analytics, whose activities rely on well-organised and clean data, which are expected to be achieved throughout the management lifecycle \cite{RN85}. Data organisation, cleansing, and structuring are crucial processes that pave the way for analysis and knowledge extraction. This section summarises the state-of-the-art data management approaches and narrows them down to practices in energy systems. To understand the position of data management approaches in both research and industry, a summary of what is currently needed, given the organisational and technological context, is also provided.
	
	\subsubsection{Common Data Management Practices}
	\label{subsec:common_data_mgt_practice}
	Traditionally, data has been managed in a row-and-column format, commonly referred to as a Relational Database. Relational Database Management Systems (RDBMS) have been popular among various computing platforms, and for years, they have been a popular and reliable choice \cite{RN63, RN144}. Their strength relies on strict rules to enforce data structure and formats. RDBMS use Atomicity, Consistency, Isolation, and Durability (ACID) properties to achieve reliable transaction processing and data integrity. 
	
	With their structure, they can easily facilitate data acquisition, storage, and extraction. Due to their strict rules, they are known for having structured and clean data that is almost ready for use. However, with the current nature of generated data, these strict rules have potential limitations towards the performance of digital systems in terms of schema and scalability. This forces a shift to NoSQL databases — flexible and scalable database systems and distributed databases \cite{RN161, RN152}. These new approaches are commonly referred to as modern data management platforms \cite{ma2024enhancing}.
	
	NoSQL databases store data in a flexible data model, outperforming their counterpart, RDBMS, for managing unstructured data. NoSQL databases are designed to expand horizontally, providing the ability to handle massive volumes of data and accommodate the high velocity and variety of big data sources. This makes them the best choice for voluminous and rapidly evolving data, as is often the case in energy systems. Similarly, distributed databases have also become a popular choice among the technical communities \cite{RN94, RN95, RN96}. These systems, among many other advantages, also offer transparent management of distributed and replicated data, reliability through distributed transactions, scalability, and improved performance \cite{RN94, RN95}. However, their implementation also raises challenges related to data control, distributed database design, query processing, and data integration, among others \cite{RN95}. 
	
	In use cases where data are characterised as big data, most opt to manage data in large cloud deployments \cite{RN221, RN154, RN72, RN235, RN213, RN106, RN251, RN376, RN254}. However, the cloud can also lead to vendor lock-in, data dissipation, cost-racking, and security challenges \cite{RN72}. Usually, due to the amount of central coordination necessary for making big data viable, this is mediated through a central authority that controls access and exchange of data on its network. This leads to a looping challenge for central data management \cite{gunther2017debating}. 
	
	Generally, data management research spans from addressing the challenges of streaming data to managing the heterogeneous nature of data that is so sparse across nodes \cite{RN148}. It can also be looked at in terms of technologies in use or application domain. Concerns usually vary among different use cases. Recently, blockchain technology has been taking charge of addressing problems that demand decentralised approaches, among other technologies. For instance, Amiri M.J. et al. \cite{RN146} explored consensus protocols used in modern large-scale data management systems to enhance fault tolerance in distributed systems. This work had two motives: first, to realise the benefits of decentralised databases, and second, to realise the benefits of blockchain technology. A combination that allowed authors to explore a consensus algorithm for node management in decentralised setups. Similar implementations also appear in \cite{RN255}, where authors proposed large-scale data management using a permissioned blockchain. However, ensuring data quality and achieving seamless data integration across systems remain significant challenges in these approaches. 
	
	On the other hand, security is another critical aspect of data management, particularly with the growing use of cloud and blockchain technologies. With rising security challenges, El Abbadi \cite{RN147} highlights the need for secure and trustworthy data management in cloud and blockchain environments. Similarly, research notes that current solutions face significant scalability and performance challenges when implemented in large-scale data repositories. Moreover, trust issues arise when data is hosted in cloud environments, as users often have a limited understanding of the underlying infrastructure, leading to concerns about data security and reliability. Several other new database designs aim to address the natural tension between performance, fault tolerance, and trustworthiness, which remain open questions for the approaches discussed. 
	
	Data management in IoT settings is another trend that has gained attention among the research community \cite{RN246, RN202, hajjaji2021big}. As highlighted before, this is highly influenced by the amount of data generated and collected by these systems.
	
	Some implementations have already been dominant in decentralised data management approaches using Hadoop frameworks \cite{RN199, RN206} and Google File System (GFS) \cite{RN173, pan2024navigating}. These systems are gaining popularity due to their ability to accommodate the data management life cycle. Naturally, they are developed to support the storage and processing of large datasets with a simplified programming model. Both systems store data in file formats, which introduces challenges related to data updates, managing schema changes, and data integration. Aspects of data storage technologies based on these data management technologies will be presented in Section \ref{subsec:data_storage}.
	
	As noticed, approaches for data management are directed towards distributed data management. Distributed databases provide seamless access to database systems. They are made of multiple computers where datasets are distributed across each computer that makes up the database cluster \cite{RN94}. Some examples include Google Spanner, Azure Cosmos and some data warehouses. The motivation underlying the use of distributed databases varies between businesses; generally, a distributed database approach could be applied when the data cannot be accommodated into a single computer system.
	In some cases, when computation related to managed data takes a very long time, a distributed approach can also be applied. Other motivating reasons include the need for resilience, fault tolerance, data locality, access control and flexibility—advantages offered by distributed databases \cite{RN94, RN95, RN92}. However, these approaches are implemented with a few concerns; data integration and consistency are still potential research questions to explore \cite{RN127, RN95, RN94}.
	
	These approaches would also contribute highly to the effectiveness of data for analytics. One reason behind this argument is that distributed databases enable the collection of data from multiple data points. This increases the volume of data and its resolution for application in analytics. Furthermore, data managed in a distributed fashion ensures the fastest response time for distributed queries \cite{RN256}. Traditional distributed data management approaches focus on collecting data in a distributed fashion and sharing it with a data centre (usually in the cloud) for processing and analytics. However, transferring large datasets to a single data centre may be impractical due to bandwidth limitations, communication latency,, time costs, and data privacy concerns.

	\subsubsection{Data Management Practices in Energy Systems}
	\label{subsubsec:dm_energy}
	When it comes to energy systems, the data management practices are not clearly defined. In most cases, each energy custodian has its own data acquisition, storage, and processing infrastructure. Hence, practices vary from one Distribution System Operator (DSO) to another, one Transmission System Operator (TSO) to another, and among other stakeholders. The defined database architectures are determined according to the specific needs of individual stakeholders, making it difficult to directly reuse the data for different purposes, especially in analytics \cite{RN135, RN277}. Data sharing and integration have become even more challenging due to the monopolistic nature of the energy sector \cite{RN124}. This section explores recent data management practices in the energy domain.
	
	Since 2020, most practices have been leveraging the benefits provided by cloud and data centre services, primarily to achieve improved performance, efficiency, and load balancing in smart grids. This is highly influenced by the newly generated data spanning from various sources. Undoubtedly, research focusing on Machine Learning in energy management systems \cite{akram2024smart}, advanced data-driven decision-making \cite{RN421}, and data acquisition methods \cite{RN422} has a significant influence, primarily aimed at improving energy efficiency. 
	
	IoT and edge computing are used at the data acquisition stage to enable real-time data collection and processing at the source, reducing latency and bandwidth usage during data communication \cite{RN376, RN276}. IoT has supported distributed data collection, which is crucial for renewable energy and demand-side monitoring \cite{RN424}. 
	
	Due to the ever-increasing amount of data in power systems, management approaches are also considering a shift from relational databases to NoSQL \cite{RN425, RN426}. Big data management tools such as Hadoop, MapReduce \cite{RN171}, HDFS and Hops File System (HopsFS), Apache Spark \cite{RN432} and custom tools like “SmartSantander”\cite{RN427}, “SCOPE”\cite{RN428}, “FIWARE”\cite{RN429} have been adopted to facilitate advancements in smart cities. In addition to smart grid data management, tools such as “SealedGRID” \cite{RN430} and anomaly detection in big data approaches \cite{RN431} have also been proposed, utilising blockchain technology. 
	
	Given the identified challenges, Zainab et al. \cite{RN424} proposed a novel big data management architecture that covers data collection, storage, transfer, and mining. The authors suggest using SCADA, Advanced Metering Infrastructure (AMI), smart meters and sensors for data collection and upload to the cloud. Apache Hadoop is also proposed for data storage, as it can integrate cloud storage and Hadoop using HDFS \cite{RN213, RN335, RN393}. Several data mining tools, such as Apache Spark \cite{RN199, RN455, RN391}, Apache Hive \cite{RN199, RN383}, and Cassandra \cite{RN428, RN113}, have also been proposed to enhance Machine Learning applications in energy systems. 
	
	The introduction of blockchain technology in energy systems has introduced decentralised coordination, targeting energy data metering, tamper-proof registration, and smart contracts \cite{RN423}. This development addresses challenges in peer-to-peer energy trading and the operation of virtual power plants. With a few customisations, it could also benefit data storage mechanisms.
	
	The disparities in system architectures, database designs, database management systems (DBMS), and data policies pose another layer of hurdles toward success in data management and analytics in energy systems. Data uncertainty is another challenge due to its complex nature because sources are dispersed and distributed. These uncertainties span between uncertain data mining and imprecise data querying \cite{RN58}. It also raises concerns about data quality, as most of the collected datasets are often incomplete, inconsistent, and uncorrected. These challenges require a range of data preprocessing technologies to enhance data quality. Since most big data are currently managed in multiple distributed grids, real-time data storage security challenges are raised. Due to this, some research work proposes cryptographic algorithms to address the challenge. However, most implementations come with performance degradation issues, hence demanding, again, an integrated solution with the technology stack that is meant to address most of the common problems. It is also recommended that a unified and comprehensive system standard be established, as different regulations on big data can cause conflicts between collaborators and lead to inconvenience in the smart grid \cite{RN58}.

	\subsection{What is needed in Energy Data Management}
	\label{subsec:needs_in_energy_data_mgt}
	The needs and requirements for data management vary according to specific business requirements, policy demands, regulations and other factors. Despite the differences, several regional efforts have been made to oversee progress in data management. This section provides an overview of stakeholders’ efforts to ensure that data becomes a common resource and is treated in a manner that makes it useful for improving services and enabling entities to make informed decisions. Several efforts targeting the energy sector, as well as those that combine the energy sector with other sectors, are briefly explored.
	
	\subsubsection{Data Exchange Reference Architecture}
	\label{subsubsec:dera}
\begin{figure*}
		\centering
		\includegraphics[width=.8\linewidth]{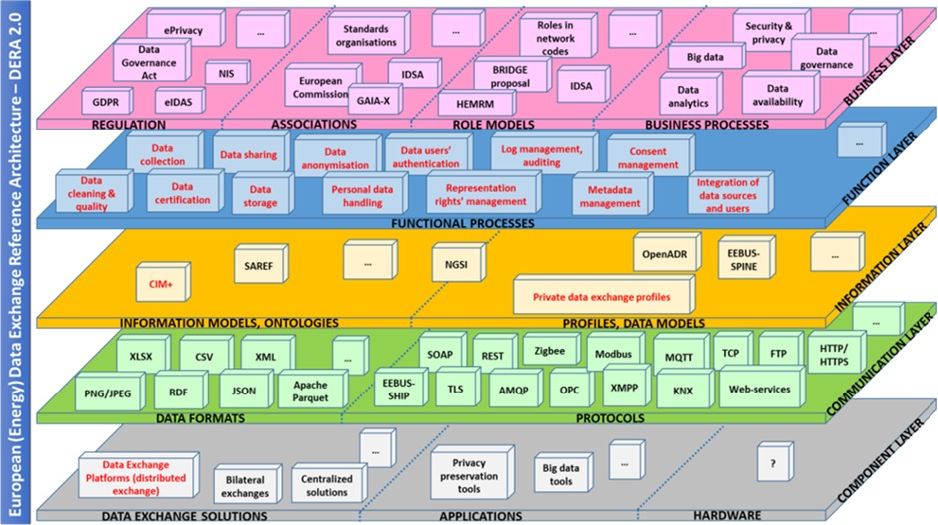}
		\caption{European Data Exchange Reference Architecture (DERA) 2.0 \cite{RN433}.}
		\label{fig:dera2.0}
	\end{figure*}

Within the context of the BRIDG  project, the data management work group focuses on guiding data exchange and processing. Their reports on “Energy data exchange reference architecture” aim to contribute to efforts toward interoperability of systems and business process agnostic data exchange techniques on a European scale for the energy domain and beyond \cite{RN434, RN264, RN433}. Its data exchange reference architecture is represented in Figure \ref{fig:dera2.0}. The structure adheres to the Smart Energy Grid Architecture Model (SGAM) to facilitate the modelling of appropriate data exchange strategies. The component layer guides data exchange strategies in three major groups: data exchange solutions, which, among many other methods, emphasise distributed data exchange strategies; then there is an application and hardware group. The communication layer emphasises protocols and data formats for data exchange strategies, insisting on open-source and widely supported platforms. Other layers are information, function, and business, which focus on information models, functional units, and governance. Its structure can be mapped to the reference architectures of other sectors, such as RAMI4.0 for industry – Reference Architecture Model Industry 4.0 \cite{RN460} and CREATE-IoT 3D RAM for health – Reference Architecture Model of CREATE-IoT \cite{RN461} project, which provides a basic interoperability vocabulary for non-energy sectors.
	
	Meanwhile, the Data Space concept also emerged in the research landscape, with concrete associations and industry clusters pushing for it from the ICT (Information and Communication Technologies) sector (such as Gaia-X, International Data Spaces Association (IDSA), Data Space Business Alliance (DBSA) \cite{RN264} etc.). These initiatives are providing new reference architectures, frameworks, and roles.  
	
\begin{figure*}
		\centering
		\includegraphics[width=.8\linewidth]{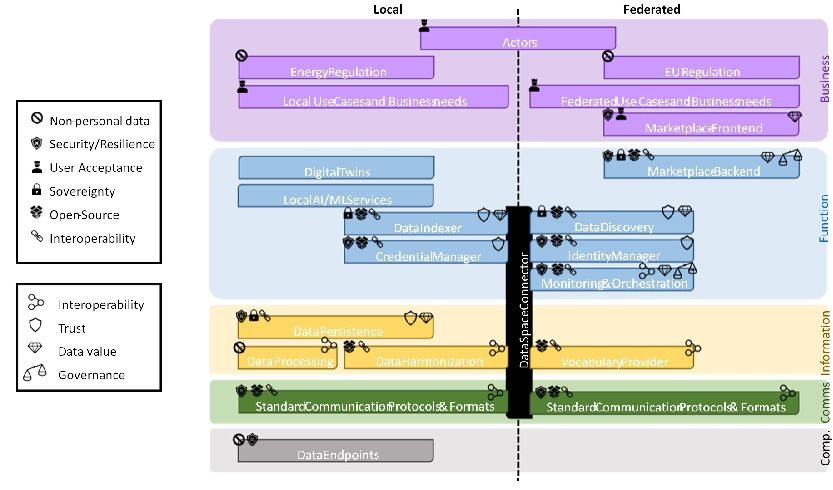}
		\caption{DERA 3.0 layered architecture and link to the DESAP \cite{RN265, RN434}.}
		\label{fig:dera3.0}
	\end{figure*}

	DERA 3.0 (Presented in Figure \ref{fig:dera3.0}) also aligns with these new inputs whilst maintaining the essence of energy-related requirements as described in previous versions of DERA, except that its simplified presentation provides implementations with a wide range of possible technical choices. Furthermore, in the European context, the European Commission also published the Digitalising the Energy System - EU Action Plan (DESAP) \cite{RN265, RN259}, which champions energy data exchange approaches and the use of ICT, among other initiatives with an emphasis on distributed data collection, storage, retrieval, and analytics.
	
	Similarly, the International Energy Agency (IEA) highlights the benefit of digitalising energy systems, for which data management is a core concern. Among other concerns, the IEA argues that the benefits of data-sharing are often overlooked, underestimated, or, in some cases, resisted, and there is a lack of incentives to invest in data and develop solutions \cite{RN260}. Privacy and data ownership are also major consumer concerns, especially as more detailed data are collected from connected devices and appliances. IEA further insists that policymakers balance privacy concerns with other objectives, including promoting innovation and the operational needs of utilities. Additionally, policymakers developing overarching data strategies should consider the energy sector as a crucial domain to explore.

	\subsubsection{Data Spaces}
	\label{subsubsec:data_space}
	Most recently, the IDSA initiated the IDSA Reference Architecture Model (IDS-RAM) — a comprehensive framework to support the creation and operation of data spaces \cite{turkmayali_2024_12663036}. Data spaces are digital environments designed for trusted sharing and management of data among various participants. They enable more efficient implementations of advanced services and solutions based on data. This is done by ensuring data sovereignty and allowing data holders to control the terms and conditions by which their data is reused. This framework encapsulates the knowledge, requirements, and findings that IDSA has accumulated over several years. The IDS-RAM is a core component of the Dataspace Protocol, which integrates key processes for data exchange, contract negotiation, and data transfer management, forming the foundation for standardised and secure data spaces. Within the framework of data spaces, data connectors play a crucial role as the primary implementation where the functionality offered by the Dataspace Protocol is realised as actual running software and services \cite{RN445}. They offer two main functionalities: (i) data exchange services and (ii) trustworthy data handling. 
	
	Janev, V. et al. \cite{RN210} explore data spaces in energy systems by analysing the challenges and requirements related to energy-related data applications. They also evaluate the use of Energy Data Ecosystems (EDEs) as data-driven infrastructures to overcome the current limitations of fragmented energy applications. EDEs are inspired by the IDSA mission. In their work, the authors focused on illustrating the applicability of EDEs and IDS reference architecture in real-world scenarios from the energy sector.
	
	A recent work by \cite{dognini2024blueprint} proposed a Common European Energy Data Space (CEEDS) - a framework that emphasises the integration of existing data platforms (including those of legacy systems) into a federated data space. As with other data space platforms, it aims to enhance data sharing, interoperability, and collaboration across the European energy sector. Several use cases are defined within the energy sector. 
	
	While providing a good guide for interoperability and data sharing, the IDSA does not offer code or implement actual technical solutions; it relies on market operators to transform specifications into workable implementations, which could lead to variability in implementation quality. The framework sets a strong foundation for trusted and interoperable data sharing. With its current design, efforts for enhancing interoperability, large-scale replications of the proposed efforts, and regulatory adoption are invited \cite{dognini2024blueprint}.
	
	The data space for energy systems efforts also identifies the following potential challenges that need to be addressed. 
	
	\begin{itemize}
		\item Fragmented data ecosystems - existing data platforms operate in isolation with limited pan-continental interconnections, introducing challenges for seamless integration into common data spaces, i.e., CEEDS (Common European Energy Data Space);
		
		\item Standardisation gaps - whilst the proposed architectures reference standards like CIM (Common Information Model), SAREF (Smart Applications REFerence), and IEC (International Electrotechnical Commission), further harmonisation is required to ensure consistent data models and ontologies across diverse systems;
		
		\item Governance complexity - the governance model for architectures like CEEDS is still under development, with uncertainties around the roles and responsibilities of the governance authority and cross-data space coordination;
		
		\item Data sharing incentives - establishing clear incentives for data sharing whilst ensuring privacy and sovereignty remains a challenge, particularly for proprietary and sensitive datasets;
		
		\item Scalability and adoption - the blueprint highlights the need for scalable solutions and widespread adoption, which may require significant investment and stakeholder alignment.
	\end{itemize}

	\subsubsection{OFGEM - Data Best Practice Guidance}
	\label{subsubsec:ofgem}
	On the other hand, in the efforts to advance quality data management, the report by OFGEM (Office of Gas and Electricity Markets) \cite{RN261, RN259} highlights data best practices covering aspects of data assets, standardisation, stakeholders, data access, data security and metadata in the energy sector. Detailed descriptions are presented in Table \ref{tab:data_best_practice}.

	{\small 
		\renewcommand{\arraystretch}{0.8} 
		\begin{table*}[t]
			\caption{Data Best Practice Guidance \cite{RN261}.}
			\label{tab:data_best_practice}
			\begin{tabular}{p{5cm}p{11cm}} 
				
				\toprule
				\textbf{Data best practice principle} & \textbf{Explanation} \\
				\midrule
				Identify the roles of stakeholders of Data Assets & Log information on data assets, data custodians, relevant data subjects, data controllers, and data processors must be identified \\
				\midrule
				Use common terms within Data Assets and Metadata & It must enable data users to search for and easily join data assets and associated metadata to data assets and metadata provided by other organisations \\
				\midrule
				Describe data accurately using industry-standard Metadata & It must make it easy for data users to work with and understand information that describes each data asset. Must, therefore, provide Metadata associated with Data Assets, and the Metadata must be made available to Data Users independently of the Data Asset itself \\
				\midrule
				Enable potential Data Users to understand the Data Assets and make them discoverable & Throughout a data asset’s lifecycle; the custodian must make available supporting information that data users require to maximise the benefits to be gained by consumers and the public. The custodian must also ensure that all potential Data Users can identify the Data Assets \\
				\midrule
				Ensure data quality maintenance and improvement & It must ensure that Data Assets are of sufficient quality to meet the requirements of their Data Users. Data Users must have the option to contest decisions regarding the definition of adequate data quality of a Data Asset \\
				\midrule
				Ensure Data Assets are interoperable with Data Assets from other data and digital services & It must enable interoperability between the data assets \\
				\midrule
				Protect Data Assets and systems in accordance with Security, Privacy and Resilience best practices & It must ensure that compliance with this guidance does not negatively impact the compliance with all relevant regulations, legislation, and Security, Privacy and Resilience (SPaR) requirements \\
				\midrule
				Store, archive, and provide access to Data Assets & It must ask stakeholders whether the data assets could create a future benefit if archived \\
				\midrule
				Treat all Data Assets, their associated Metadata and software scripts used to process Data Assets as Presumed Open & It must treat all Data Assets, their associated Metadata, and software scripts used to process Data Assets as Presumed Open and subject them to Open Data Triage \\
				\bottomrule
			\end{tabular}
		\end{table*}
	}
	
	
	Adhering to these data best-practice principles ensures that data assets are managed effectively, promoting transparency, interoperability, and quality. Organisations facilitate easier data discovery and comprehension by clearly defining stakeholder roles, standardising terminology, and providing accurate metadata. Maintaining high data quality, ensuring interoperability, and adhering to security and privacy standards further enhance data utility. Additionally, treating data assets as presumed open and subject to necessary evaluations encourages broader access and innovation, ultimately benefiting consumers and serving the public interest.
	
	\subsection{Data Storage Approaches}
	\label{subsec:data_storage}
	Despite the scarcity of works addressing data storage challenges, specifically in the energy sector, data storage practices from other fields can be borrowed, given that the use cases are similar. In many use cases that generate large datasets, their storage using traditional database architectures has proven to be inefficient. A few aspects that have been emerging are schema-on-write challenges, cost of storage, cost of proprietary storage, complexity, heterogeneous data, and integration among data sources and with other programs \cite{RN104}. 
	
	Big data storage technologies were developed to address these challenges. They are mainly used in the health and finance domains, among others. Like in general data management approaches, some advanced approaches opt for Hadoop programming modules, cloud architecture, or a combination of both. Here, we briefly summarise these implementations and highlight their limitations. Whilst storage technologies may be decoupled from DBMSs, the discussion will encompass a combination of both storage architecture and DBMSs, with the addition of software technologies as necessary. Cloud storage is also discussed, as it appears to be a common approach and offers a model for hybrid solutions.
	
	\subsubsection{Relational Databases}
	\label{subsubsec:relationalDB}
	The relational database model is one of the dominant database architectures. Since relational databases guarantee the ACID properties, they can provide a means for data storage and allow more collaboration, reliability, security, and consistency \cite{RN62}. Hence, they excel at maintaining data integrity and enforcing constraints that prevent inconsistencies and ensure accuracy. Furthermore, by running complex queries, relational databases enable users to extract specific and detailed insights, providing accurate answers to complex questions based on precise and reliable data \cite{RN471, RN63}.
	
	In energy systems, relational databases have mainly been implemented for energy management systems (EMS), SCADA systems, and grid operation systems, where structured data (e.g., energy consumption, sensor readings, and historical data) is essential. This is demonstrated in \cite{RN446}, where authors present an integrated energy accounting framework using relational database technology to facilitate detailed tracking of energy production, distribution, and consumption, ensuring transparency and accountability. 
	
	Yang et al. \cite{RN447} conducted a rare prediction study using relational databases. They explored the development of an eco-friendly DBMS by analysing and modelling energy consumption in relational database workloads. This study also introduced a novel approach that combines self-powered wireless vibration sensors (WVSs) with the Least Square Support Vector Machine (LSSVM) algorithm to create an Energy Consumption Model (ECM). The study validated the performance of self-sustaining WVSs and evaluated the accuracy of the ECM in predicting energy consumption during SQL statement execution, achieving a maximum prediction error rate of 10\%. Some other explorations combine relational databases and other storage architectures \cite{RN452}.
	
	In the context of renewable energy monitoring, Trillo-Montero et al. \cite{RN448} implemented an orderly, accessible, fast, and space-saving storage system that allows transferring to an RDBMS all data corresponding to a set of photovoltaic (PV) systems whose behaviour is to be analysed.
	
	Despite their strength, relational databases struggle to manage growing and complex datasets \cite{RN63}. In such scenarios, managing relationships is impractical. Furthermore, running complex queries in such datasets raises concerns about performance, but given the current nature of relational database design, it is also impractical \cite{RN85}.
	
\begin{figure*}
		\centering
		\includegraphics[width=.6\linewidth]{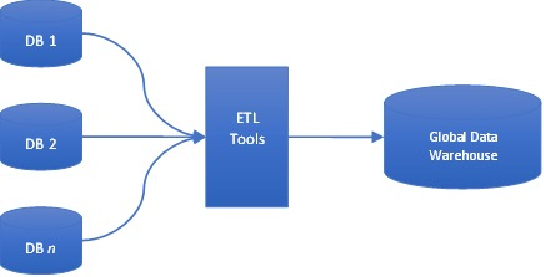}
		\caption{Data warehouse approach.}
		\label{fig:dw_approach}
	\end{figure*}

	Data warehouses were then developed to address some of the challenges that relational database systems face, including query performance and analysis. Data warehouses integrate datasets coming from multiple sources to provide efficient data analytics, reporting, and consistency. Whilst traditional databases are primarily for transactional data, data warehouses store both historical and current data. A data warehouse comprises a database, Extract, Transform and Load (ETL) layer, access tools, and metadata. In this regard, a DBMS (based on relational database architecture) remains the foundation for the data warehouse. Then, ETL extracts and prepares data, whilst access tools help users search and query, providing the context and definition for big data \cite{RN74}. A basic structure representing the data warehouse approach is presented in Figure \ref{fig:dw_approach}.
	
	Data warehouses are affected by a high rate of failures, with more than 50\% of data warehouses failing at one point — not only because of the technical challenges or complex architecture but also because of the failure of the project to meet user requirements and ever-changing business demands \cite{RN155, RN156}. Furthermore, businesses experience challenges when loading data into data warehouses, with the biggest inhibitors being legacy technology, complex data types and formats, data silos, and data access issues tied to regulatory requirements \cite{RN157}. Similar challenges are also experienced by data lakes, which, unlike data warehouses, accommodate unstructured data and embrace schema-on-read design, offering much more flexibility and scalability \cite{RN163}.
	
	In the energy sector, deploying data warehouses and data lakes has become instrumental in managing and analysing modern energy systems' vast and diverse datasets. These technologies serve distinct yet complementary roles, each tailored to specific data management and analytical requirements. Common use cases include historical data analysis, regulatory compliance and reporting, and performance monitoring \cite{RN449, RN450}. In the case of data lakes, implementations in energy systems are scarce but could benefit from the integration of diverse data sources, advanced data analytics and Machine Learning, and real-time processing \cite{RN163, RN119, RN166, RN165, RN164, RN54, RN451}. 
	
	The evolution of big data, the technological advancement, and the burst production of semi-structured and unstructured data further showed the limitation of relational data and RDBMS use and the difficulty in managing colossal data growth. The reason is that relational databases scale vertically, hence suffering from hardware constraints, i.e., the number of physical devices that can be added. Partitioning can create problems while joining tables and might lead to discrepancies \cite{RN63}. Similarly, maintaining the structure becomes challenging as the data grows and can slow the ETL process in data warehouses. Data lakes, on the other hand, face data discovery, extraction, cleaning, integration, and versioning challenges \cite{RN163, RN164, RN167}. Because of this, maintaining data quality, ensuring consistency, and realising the usability of large-scale data remains a significant challenge, often leading to data swamps — unmanaged data lakes \cite{RN166, RN165}.

	\subsubsection{NoSQL}
	\label{subsubsec:nosql}
	NoSQL databases are usually considered alternatives to relational databases due to their flexibility in design \cite{RN471}. The term NoSQL was first used in 1998 by Carlo Strozzi as the name of the file-based database he was developing; since then, it has been used for databases that omit the use of SQL \cite{RN272}. However, it was not before 2009 that it became a serious competitor to the relational database. The wide usage of these NoSQL products encouraged other companies to develop their own solutions and led to the emergence of generic NoSQL database systems. NoSQL databases are highly influenced by the need to handle huge datasets of unstructured data \cite{RN158, RN159} and the need for big data analytics \cite{RN111, RN134}. They also guarantee efficiency in supporting data aggregation for business intelligence and data mining \cite{RN160} and are further well suited for cloud computing and for storing big data \cite{RN158, RN152, RN161, RN104}.
	
	A large set of query languages can also be used with NoSQL without following a strict and predetermined schema. These databases can be designed using different models, including key-value, document, graph, or column-based models \cite{RN158, RN160}. Documents-based models are utilised primarily for the storage and administration of document-based data. Complex data formats, such as JSON, BSON, XML, and PDF, are used to store information in document-oriented databases \cite{RN116}. Key-value databases, sometimes referred to as key-value stores, use simple key-to-value methods to store data \cite{RN273}. A key is always a string (ideally unique) with an arbitrarily large field as its value. This structure makes them a straightforward option for data storage \cite{RN273}. Graphs are also gaining popularity. They present data in graph nodes and edges, which show relationships between nodes. Its structure makes it the preferred choice due to its organised structure of mapping datasets, but more importantly, traversing through huge datasets that are in graph structure is more efficient, fast and accurate \cite{RN274}. 
	
	In energy systems, NoSQL has been used in \cite{RN268, RN269, RN267, RN472} with primary use cases focusing on energy consumption data and analytics \cite{RN268, RN269, RN267}, but it is also increasingly used in smart grids \cite{RN282, RN363} and smart metering infrastructure, which generates large amounts of unstructured or semi-structured data. NoSQL’s flexibility makes it well-suited for handling these complex and diverse data types \cite{RN270, RN271}. 
	
	Despite their valuable features, NoSQL databases face several challenges, owing to their need to scale horizontally — adding nodes to distribute the database workload. NoSQL databases often don't fully support ACID properties (which is vital for structured data) because ensuring strict consistency across distributed nodes sacrifices the scalability and availability needed for horizontal scaling. This makes them inconsistent and slightly lagging in standardisation \cite{RN162}. Furthermore, scaling would remain a serious concern for these architectures when implemented in a centralised fashion because it would require vertical scaling — adding more hardware resources, in particular processing power, to a single machine \cite{RN62}. Scaling up in database systems involves adding CPU and RAM resources to increase a single server's processing speed or storage capacity to cater to growing storage requirements. Data stored in NoSQL databases raises a question of compatibility when merging data from different sources. This is similar to integration and update mechanisms. Also, handling update rates with ever-increasing amounts of data (velocity) is still a puzzle \cite{RN104}. On the other hand, most current NoSQL databases do not address the challenges related to data sharding, leaving it as a research question. Despite their ability to address a general aspect of scalability, there are still concerns when scalability involves changes in the database schema.
	
	\subsubsection{Hadoop}
	\label{subsubsec:hadoop}
	Hadoop is a programming framework designed to store and process large datasets across clusters of computers, effectively scaling from a single machine to thousands \cite{RN180}. It achieves this by dividing workloads into smaller tasks that can be executed concurrently. This addresses the challenges of vertical scaling introduced by other approaches. The Hadoop ecosystem comprises four core modules: HDFS, which provides high-throughput access to application data by allowing nodes to process data stored locally \cite{RN181}; Yet Another Resource Negotiator (YARN) which is responsible for resource management and job scheduling within the cluster \cite{RN170}; MapReduce, a programming model that processes large-scale data by distributing tasks across multiple nodes for concurrent execution \cite{RN171}; and Hadoop Common, which offers essential libraries and utilities shared among the other modules. Beyond these primary modules, the Hadoop ecosystem has expanded to include tools such as Apache Pig, Apache Hive, Apache HBase, Apache Spark, Apache Zeppelin and Presto, each enhancing various aspects of big data collection, storage, processing, analysis, and overall management.
	
	Hadoop has been a potential storage architecture in the energy domain. A few common use cases include smart grid management \cite{RN423, RN430, RN114, RN424}. In this case, it facilitates the storage and analysis of large datasets generated by smart grids, enhancing grid reliability and efficiency \cite{RN424, RN453, RN454}. Some explorations focus on the integration of renewable energy sources \cite{RN201, RN233} and energy consumption monitoring \cite{RN234, RN452, RN268, RN270, RN403, RN241}.
	
	Despite its massive deployment across several domains, Hadoop can’t handle modern Information Technology (IT) systems in data velocity, scalability, and Machine Learning algorithms \cite{RN455}. It is also not very efficient as it cannot produce output in real-time with low latency \cite{RN424}. Mechanisms for Hadoop operation demand that the so-called master node (Hmaster) controls the worker nodes (running mapper and reducer tasks), giving a sense of centralised control. Integration of these platforms with existing systems introduces another layer of complexity in deployment \cite{RN459}. It is also not ideal for real-time data querying since it has been designed for batch processing. This gap underscores the need to integrate real-time streaming technologies like Apache Kafka. Similarly, complex queries involving complex relationships cannot be run. HDFS, the storage infrastructure for Hadoop, is designed to work properly with a small number of large files rather than a large number of small files.
	
	\subsubsection{Blockchain-based Data Storage}
	\label{subsub:blockchain}
	Blockchain is a decentralised and distributed digital ledger technology that records transactions across multiple computers in a way that ensures the data is secure, transparent, and immutable. Each record, or block, contains a cryptographic hash of the previous block, a timestamp, and transaction data. This structure makes it nearly impossible to alter any information without altering all subsequent blocks, which requires consensus from the network participants. 
	
	Blockchain is widely used for cryptocurrencies like Bitcoin and Ethereum, but its applications extend to various fields such as supply chain management, healthcare, power systems and finance due to its ability to provide secure and transparent data management. Given its decentralised nature, it represents a unique opportunity for decentralised data management \cite{RN255}.
	
	Blockchain treats connected nodes in the network as peers, giving them full autonomy. Reflecting this into data management can significantly enhance big data applications by addressing several key challenges and providing various benefits. Blockchain's decentralised and encrypted nature ensures that data is secure and private. The immutability of blockchain ensures that it cannot be altered once data is recorded. Blockchain enables secure and efficient data sharing among multiple stakeholders. By using blockchain, data scientists can access high-quality, structured, and complete data.
	
	Several implementations exist, including \cite{RN255}, where authors proposed large-scale data management using a permissioned blockchain. The main focus of this work was to address four major challenges with blockchain. First, confidentiality: authors introduced a permissioned blockchain system that supports both internal and cross-enterprise transactions of collaborating enterprises. Second, verifiability: the authors introduced a blockchain-based multi-platform crowd-working system that enforces global constraints on distributed independent entities. This is done by ensuring that privacy is preserved using lightweight and anonymous tokens whilst transparency is achieved using a permissioned blockchain shared across multiple platforms. Third, performance: the authors introduced a new paradigm for permissioned blockchains to support distributed applications that execute concurrently. Fourth, scalability: the authors introduced a permissioned blockchain system that improves scalability by clustering (partitioning) the nodes and assigning different data shards to different clusters where each data shard is replicated on the nodes of a cluster.
	
	The authors also addressed the challenge of fault tolerance — by introducing a hybrid State Machine Replication protocol that uses the knowledge of where crashes and malicious failures may occur in a hybrid environment to improve overall performance. However, this implementation and others that are based on blockchain \cite{RN246, RN117, RN234} do not address the challenges of data quality and data integration at a large scale.
	
	On the other hand, security is another critical aspect of data management, particularly with the growing use of cloud and blockchain environments, as emphasised by \cite{RN147}. Similarly, current solutions face significant scalability and performance challenges when implemented in large-scale data repositories \cite{RN213, RN256}. Moreover, trust issues arise when data is hosted in cloud environments, as users often have limited insight related to the underlying infrastructure, leading to concerns about data security and reliability.
	
	In the context of energy systems, blockchain can improve the management of energy systems by enabling secure and transparent energy trading, decentralised energy generation, and efficient data management \cite{RN234}. This is currently being implemented mainly in the context of smart grid management. In some research \cite{RN320, apeh2025enhancing}, blockchain is explored for its role in facilitating transparency, providing immutability and trust mechanisms for secure energy management, and integrating AI and cloud. Another work in \cite{RN431} focused on anomaly detection by providing a tamper-resistant and distributed ledger system. The proposed framework leverages blockchain to support distributed and collaborative anomaly detection. Gagliardelli et al. \cite{RN452} promote good practices in local energy communities by leveraging big data platforms to collect and analyse data, and utilising blockchain for energy tokenisation with smart contracts. 
	
	Yet, blockchain still experiences several limitations. Permissioned blockchain systems, for instance, face serious challenges in terms of confidentiality, performance, scalability, fault tolerance, and verifiability. Nevertheless, it is important to note that some of these challenges have been addressed in \cite{RN255}. Maintaining confidentiality is difficult as a single ledger stores all transactions, exposing internal enterprise data. Scalability and performance are also limited by the requirement for all nodes to process every transaction, with proposed solutions such as sharding struggling to handle cross-shard transactions efficiently. Scalability limitations persist because every node must store a complete copy of the dataset, resulting in high redundancy, latency, and storage costs. Their sequential data structure and reliance on consensus mechanisms limit the flexibility of real-time analytics and querying. Fault tolerance issues persist in hybrid environments that include both trusted and untrusted nodes, resulting in performance inefficiencies. Addressing these limitations requires advanced protocols to improve these aspects in blockchain-based data management systems.

	\subsubsection{Peer-to-Peer Data Storage}
	\label{subsubsec:p2p}
	Peer-to-peer (P2P) network systems initially gained popularity in the context of content and file sharing. Initial implementations of distributed DBMSs, focusing on P2P architectures, demonstrated no differences in the functionalities of each site within the system. Modern P2P architectures offer three major improvements: massive distribution, the heterogeneous nature of sites and their autonomy, and the volatility of systems \cite{RN95}.
	
	In satisfying the new demands, modern P2P systems aim to achieve autonomy — join and leave at will, query cost, efficiency — a high number of queries processed in a given time interval, quality of service — user-perceived system efficiencies such as completeness of query results, data consistency and availability, query response time, fault tolerance, and security. 
	
	The strength of P2P data management relies on its flexible design choices, which match varied user requirements. Major P2P network overlays are classified into pure P2P and hybrid P2P. In pure P2P overlays, there are no differences between nodes — all are equal \cite{RN55}. In hybrid overlays (also known as superpeer systems), some nodes are tasked with management roles to oversee other nodes or are given special tasks to perform \cite{RN236}. The implementation of pure P2P may be based on either structured or unstructured overlays. With an unstructured P2P overlay, there are no restrictions on data placement, whilst structured overlays follow a principled way of organising peers in an overlay, providing a distinct advantage for scaling \cite{RN83}. Structured P2P is also known as Distributed Hash Table (DHT), and its indexing and data location mechanism facilitate content lookup and retrieval in an overlay \cite{RN84}.
	
	The implementations of P2P data management are scarce, and based on the current search, none are available for energy systems. Among the existing works, the authors of \cite{RN457} present a logical formalisation of P2P data integration systems based on classical first-order logic and an epistemic approach. Unfortunately, these methods face significant challenges. They treat the entire P2P system as a single, unified logical entity, concealing each peer's individual role and structure. This unified approach overlooks the diverse and autonomous nature of different peers within the system. In complex P2P networks, determining whether a query can be answered (i.e., decidability) becomes impossible, even if each peer's structure is simple. Additionally, the interconnected nature of peers means that constraints or rules from one peer can unintentionally affect others, leading to unintended consequences and complicating the system's overall behaviour.
	
	Another work by Akbarinia and Martins \cite{RN458} utilised DHT lookup, addressing challenges identified in the previous P2P approach. The authors present Atlas P2P Architecture (APPA) — a data management system for large-scale P2P and Grid applications using P2P simulation environments like JXTA, Chord, and CAN. APPA focuses on two main features, data availability and data discovery, which are two main requirements for supporting the Open Grid Services Architecture (OGSA)-P2P. Data availability is ensured through replication using multiple hash functions and timestamping. The Persistent Data Management (PDM) service replicates data across several nodes and uses logical timestamps to maintain the consistency of replicas. Data discovery is facilitated by query processing, which supports schema-based queries and further considers data replication. The query processing involves four main phases: query reformulation, query matching, query optimisation, and query decomposition and execution. It also supports Top-k queries to limit the number of results returned to the user, improving efficiency and user experience. Maintaining mutual consistency of replicated data after updates, especially when nodes leave the network or updates occur concurrently, is difficult in this architecture.
	
	Another work in the context of P2P focused on enhancing data privacy \cite{RN462}. The authors proposed a model that integrates purpose-based access control, trust management, and cryptographic techniques to ensure data privacy, specifically for P2P setups. Although this system covers the P2P data management domain, its focus was not to address challenges related to data storage and integration. However, it remains a recommended guide for privacy-preserving data management in P2P networks.
	
	One of the recent works \cite{RN463} proposed Hydra — a P2P decentralised storage system that enables decentralised and reliable data publication capabilities. Hydra enables collaborating organisations to create a loosely interconnected and federated storage overlay atop community-provided storage servers. Whilst addressing the solution for storage systems, its implementation hardly focuses on storing large, complex files. The authors use a name-based integration approach. Each piece of data is assigned a semantic name, which is used for all operations, including publication, access, replication, and security. Focusing on the names and metadata associated with each dataset rather than the underlying data structure allows for flexible and community-specific customisation of data naming, making it easier to manage and retrieve data. However, it loses the benefit of possible data integration that other database management services offer. Generally, Hydra enables decentralised file storage and retrieval, not datasets enabled by DBMSs.
	
	P2P systems have traditionally been associated with file sharing; they hold significant potential for energy systems. They can enable decentralised data sharing among grid operators, reducing bottlenecks in central repositories. Notably, using P2P architectures for data exchange (using DBMSs) in fully decentralised settings remains largely unexplored. These systems inherently address scalability challenges by supporting many peers, enhancing availability, and facilitating self-organisation. In parallel, it brings the DBMSs into a decentralised environment, promising an even more efficient system. Such characteristics make P2P architectures promising for applications beyond content sharing. Still, the few works that explored data management and fusion in P2P setups have failed to ensure data consistency is achieved across replicas in highly dynamic setups. Furthermore, data integration in P2P setup by benefiting from query processing approaches is still a research concern.
	
	\subsubsection{Google File System}
	\label{subsubsec:gfs}
	GFS aims to meet the demands of large-scale data processing. Its design is highly guided by other distributed data management implementations, focusing on major goals such as scalability, availability, fault tolerance and reliability. The GFS design was driven by key observations of Google’s application workloads and technological environment. These choices departed from the then-existing systems because they could not fulfil the computational demands of the company. Furthermore, GFS was designed as a distributed file system to be run in clusters of up to thousands of machines, coming with a programming interface that helps in the abstraction of management and distribution issues during development \cite{RN172}. 
	
	Just as in Hadoop, component failure is treated as a norm, files are considered to be huge, and most files are mutated by appending new data rather than overwriting existing data to realise GFS’s functionality \cite{RN172, RN173}. To the best of the knowledge established in this work, the implementation of GFS in energy systems is non-existent.
	
	Despite its maturity, several limitations still exist. GFS was mainly developed for Google, so its goal was to achieve its objectives, but it cannot always fit into targeted use cases. GFS also implements a single master node to oversee the system’s namespace and operations. This draws attention to a single point of failure. Metadata scalability is also limited with the increasing number of files. GFS adopts an eventual consistency model, meaning that changes to the file system may not be immediately visible to all clients. This can lead to temporary inconsistencies in file states, which may not be suitable for applications requiring strong consistency guarantees \cite{RN172}.
	
	\subsubsection{Cloud as a Service}
	\label{subsubsec:cloud}
	To address the limitations of traditional DBMSs and the complexities of managing data lakes, many companies leverage cloud providers like Amazon Web Services (AWS), Google Cloud Platform (GCP), IBM, and Microsoft Azure \cite{RN73}. These providers offer global accessibility and massive scalability, enabling businesses to manage and process data more effectively \cite{RN73}. The cloudification of traditional DBMSs has become a widely adopted practice in research and industry, particularly as organisations struggle with big data management challenges. Distributed database architectures have evolved to harness cloud capabilities, addressing scalability, availability, and performance issues.
	
	Cloud services are available in various paradigms, including Infrastructure as a Service (IaaS), Platform as a Service (PaaS), Software as a Service (SaaS), and Database as a Service (DBaaS), allowing users to select solutions tailored to their needs.  Data-as-a-service (DaaS) focuses on data aggregation and management through web services like RESTful APIs. Meanwhile, Database-as-a-Service (DBaaS) provides managed databases, supporting relational and non-relational databases, often distributed across cloud environments. In most cases, storage-as-a-service (STaaS) encompasses DaaS and DBaaS, providing comprehensive storage solutions. 
	
	An emerging cloud service model is Big Data-as-a-Service (BDaaS), which facilitates the migration of traditional big data applications (e.g., Hadoop) to the cloud \cite{siddiqui2025redefining}. BDaaS typically integrates three key components: (i) Infrastructure as a Service (IaaS) — to provide the underlying computational and storage resources; (ii) Storage-as-a-Service (STaaS) — a subset of PaaS that dynamically scales data storage and management; and (iii) Data Management Services — to address tasks such as data placement and replica management.
	
	In energy systems, cloud storage implementation appears to become a common go-to solution. This is evidenced by several works, including \cite{RN235, RN435}, which used the cloud for smart-grid management and energy management systems touching on data fusion, analysis, storage, and security. Other uses include the exploration of renewable energy integration by managing data from diverse inputs, such as solar panels and wind turbines \cite{RN436,zhang2025verifiable}. Some practices also explore predictive maintenance \cite{RN277, RN438} by analysing data collected via cloud platforms, enabling energy companies to predict equipment failures and schedule maintenance proactively, thereby reducing downtime and operational costs \cite{RN439}.
	
	Cloud computing enables the implementation of demand response strategies, allowing utilities to adjust energy supply based on real-time consumption data, thereby balancing load and preventing grid overloads. The role of cloud computing in power systems, including its drivers, challenges, and real-world use cases, has also been explored in scholarly works \cite{RN440, RN441, RN442, wang2025cloud}.
	
	Despite these benefits, the cloud, in its entirety, can lead to vendor lock-in, data dissipation, cost racking, and security challenges \cite{RN73, RN72, RN132}. Migrating data between cloud platforms is often impractical due to architectural differences, making it difficult for users to transfer data across cloud storage services \cite{RN132}. To mitigate these issues, some businesses adopt the polynimbus approach, which utilises multiple clouds simultaneously. In contrast, others employ a hybrid cloud system to address the challenge of relying on a single cloud provider \cite{RN71}. These approaches may increase the complexities of data management.

	\subsection{Data Integration}
	\label{subsec:data_integration}
	Data integration involves methods for joining data that are typically sourced from different sources. Considering the nature of modern data, data integration is regarded as one of the hottest challenges in research and industry. This is mainly because data is massively generated in a distributed nature. Data integration becomes a vital data management attribute because it ensures that shared data are complete, accurate and of high quality. Here, we explore methods for data integration and fusion, along with their associated challenges. We focus on innovative approaches in the computer science domain and apply them to the energy domain. It should also be noted that the explored methods only target distributed sources, as traditional approaches with centralised data are mature and effective.

	\subsubsection{Approaches for Data Integration}
	\label{subsubsec:approaches_data_integration}
	Data integration efforts can be traced back to the development of the Human Image Database (HID). HID is an extensible database management system developed to handle large and diverse datasets collected in clinical imaging communities.  One of the functions implemented was the ability to run distributed queries and integrate the distributed sources. The biomedical experts initiated later efforts to define data dictionaries and vocabularies \cite{RN470}.
	
	A work by Azza et al. \cite{RN464, RN475} explores the integration of large-scale data processing systems like Google MapReduce and Apache Hadoop with traditional parallel DBMS such as Greenplum and Vertica. In this work, integration is done by translating each SQL query to MapReduce jobs for each node, hence enabling data fetching in the targeted repository. They developed HadoopDB, a hybrid system that combines Hadoop’s scalability and fault tolerance with the high performance and efficiency of parallel DBMSs. By doing so, they highlight the benefits of combining the strengths of large-scale data processing systems and parallel DBMSs, leading to significant performance improvements and new capabilities in data processing frameworks. 
	
	Lv Z et al. \cite{RN127} explore data fusion and data cleaning systems for smart grids' big data. The system integrates multi-source heterogeneous grid data into a unified format, making it easier for computers to recognise and process data. It involves converting text and database files into a standardised or, as it is named, unified format (CSV). Data are assumed to be stored in a distributed fashion, whilst the data fusion process ensures that data from various sources and formats is standardised, making it easier to perform subsequent data mining and analysis tasks.
	
	Data fusion for power systems has also been explored in \cite{RN438}. The proposed data fusion method is based on a decentralised architecture; it involves integrating high-dimensional data from multiple regional systems to monitor oscillatory behaviour in power systems. Data from various control areas or utilities, each with its network of sensors, are collected at regional Phasor Data Concentrators and horizontally concatenated into a multi-block representation. This representation is analysed using Multi-view diffusion maps, multi-block Principal Component Analysis, and other tensor-based methods to capture within-block variances and between-block covariances. The process ensures consistency in units, dimensions, and magnitude through scaling techniques and emphasises sensor placement and clustering techniques to capture system dynamics. This approach facilitates the identification of dynamic trends and fault-dependent mode shapes directly from transient stability simulations, enhancing the accuracy and efficiency of power system health monitoring. Hence, concurrently addressing issues such as noise, missing data, and computational complexity. The authors suggest that next-generation data fusion models must effectively integrate heterogeneous data to enhance situational awareness.
	
	Similar work explored data sharing in energy systems by introducing the commodity attribute of data assets and explaining the bottlenecks of data trading \cite{RN124}. Two critical issues are reviewed: (i) data right confirmation and (ii) privacy protection, which provide a fundamental guarantee for credible data openness. Despite not technically presenting a solution for data integration, this work can be considered a guide to data sharing among entities, including federated data custodians. Data sharing is covered in this review because it is necessary within data integration frameworks.
	
	In supporting efforts towards data sharing, Hutterer and Krumay \cite{RN467} emphasised the role of data sharing in staying competitive in the market by leveraging platforms like data spaces. Their work identifies two main dimensions - technical and management — each with several sub-dimensions that challenge integrating heterogeneous data sources in data spaces. The technical dimension includes integration, indexing, querying, user feedback, and security, whilst the management dimension covers organisational and cross-organisational implementation. It is emphasised further that, whilst technical challenges can be addressed, organisational and social issues remain significant barriers to adopting data spaces. Further research on the relationship between data spaces and data ecosystems, focusing on trust and data sovereignty, is recommended. 
	
	Blockchain technology is equally mentioned in efforts toward data integration. One of these is presented in \cite{RN405}, highlighting prevailing challenges on interoperability and schema design, data indexing and supply-to-demand matching, copyright, and data quality.
	A recent exploration by Walha et al. \cite{RN465} focused on ETL (Extract, Transform, Load) to explore a data integration approach based on traditional and big data systems. Within ETL, the extraction refers to data being collected from various sources to gather all relevant data needed for analysis.  Transformation applies multiple pre-processing techniques such as data cleaning, data normalisation/scaling, missing data imputation, etc. Business rules are then applied. Lastly, loading focuses on uploading transformed data to a target system. The authors emphasised the integration of distributed computing frameworks like Hadoop and MapReduce into ETL processes. ELT (Extract, Load, Transform) is equally important with its focus on moving towards cloud platforms and handling semi-structured and unstructured data. While these technologies enhance data collection, storage, and processing capabilities, the authors further emphasised the need for more generic and customisable ETL design models to ensure reusability and flexibility in big data contexts.
	
	Popular approaches for data integration can be grouped into several major groups summarised below:
	
	\begin{enumerate}
		\item Cloud-based integration and data fusion — as observed in the reviewed articles, most storage approaches are shifting toward cloud solutions. Hence, practices have increasingly focused on cloud computing for data integration \cite{RN295, RN204, RN207, RN228, heidari2024reliable, RN221, RN72, RN235, RN285}, leveraging its scalability to handle massive datasets. Common solutions have explored data fusion techniques to integrate heterogeneous data sources, improving the reliability and robustness of analytics. This approach aims to minimise uncertainty and enhance the quality of big data analysis by combining data modalities effectively.
		
		\item ETL/ELT Data integration — ETL data integration is common and widely used in traditional databases \cite{RN123, RN119, RN465, RN186}. Its methods have been implemented in parallel with other advanced storage technologies like Hadoop \cite{RN465};
		
		\item Data spaces — the rise of data spaces cannot stay unnoticed \cite{RN221, RN264, turkmayali_2024_12663036, RN210, RN467, RN377, RN381, dognini2024blueprint}. They offer an abstraction in data management that addresses some limitations of traditional data integration systems. They also provide base functionality over all data sources, regardless of their level of integration, and allow for incremental improvement as needed. This pay-as-you-go approach reduces the initial effort required to set up a data integration system and allows for gradual enhancement based on user requirements.
	\end{enumerate}
	
	As shown, most innovative approaches do not particularly address the energy sector. Instead, it has been common for energy system practices to be adopted from other domains, particularly computer science. Data integration approaches are not an exception. Nevertheless, there have been efforts in the direction of the energy domain.
	
	\subsubsection{Limits}
	\label{subsubsec:limitation}
	One significant challenge common to all the proposed approaches is the volume and complexity of the data. This complicates traditional data integration approaches, such as ETL/ELT and MapReduce, as well as cloud solutions. The scale of data necessitates scalable storage and processing solutions, such as Hadoop, but these systems are reported to struggle with the processing speeds required for timely decision-making \cite{RN124}. Furthermore, energy systems often generate a large amount of redundant data due to their stable operation and concrete data acquisition, leading to low data value density. This redundancy can overwhelm traditional ETL processes, which are not inherently optimised for filtering out less valuable data. 
	
	In the cloud, despite their scalability, the proposed approaches face challenges related to data redundancy and the need for high processing speeds to support real-time analytics. In particular, the need for high-speed data processing to support real-time decision-making in energy systems poses a challenge, as conventional systems might not meet the speed requirements \cite{RN124}. Most cloud services focus on providing scalable and reliable storage solutions, leaving the responsibility for ensuring data quality — such as cleaning, validation, and deduplication — to the user or application layer. Whilst some advanced cloud services include tools for improving data quality, these are typically not their primary focus.
	
	With MapReduce-based systems like Hadoop, performance inefficiencies arise, particularly with complex SQL queries. The forced materialisation of intermediate data and limited support for various join operations can slow down processes, making these systems less efficient for certain analytical tasks \cite{RN465}. In our review, we did not find approaches that support query processing while addressing efficient data integration challenges.
	
	Traditional database systems and even hybrid solutions, such as HadoopDB, require extensive initial data preparation, including modelling, schematization, and tuning. This preparation phase is time-consuming and often requires substantial human effort, which can delay the integration process. Furthermore, initial versions of some systems did not optimise schema generation, leading to suboptimal performance, especially with nested or semi-structured data \cite{RN464}. While HadoopDB and similar systems have shown promise in handling structured and semi-structured data, they initially lacked the capability to efficiently manage unstructured data, which is increasingly common in big data environments. This limitation can hinder comprehensive data integration, as it requires consideration of all forms of data.
	
	An aspect of fault tolerance and scalability in this ever-scaling age is vital. Mid-query fault tolerance is another limitation in traditional parallel database systems, and it is becoming increasingly critical as data volume and system scale increase. Failures can disrupt operations, and recovery can be a complex process. Similarly, the decentralised data fusion model, although innovative, struggles with scalability when dealing with very large or highly complex datasets, which impacts real-time analysis capabilities \cite{RN66}.
	
	Sharing data across different entities, particularly in energy systems, raises significant concerns regarding privacy and security. The decentralised approach of data fusion requires mechanisms to protect data during transmission and integration, which adds layers of complexity to the system design \cite{RN66}.
	
	Current integration approaches do not accommodate real-time data integration. As IoT and Industry 4.0 gain momentum, the demand for real-time data integration has increased. Traditional ETL processes are not inherently designed for streaming data. Hence, the adoption of technologies like edge computing and others similar to it to reduce latency becomes vital. However, integrating these technologies into existing frameworks remains a challenge, especially in ensuring seamless real-time data fusion and analysis.
	
	Apart from data volume, data heterogeneity remains at the core of the problem in data integration approaches. This is because integrating heterogeneous data from distributed sources would require a new computing paradigm. For instance, systems with multiple databases deployed would require query translation later to fit into the DBMS of each local node. An easy approach is to enforce a common schema across the system. Whilst this might be effective in some use cases, in others, it is impractical and ineffective.

	\subsection{Summary}
	\label{subsec:summary}
\begin{figure*}
		\centering
		\includegraphics[width=.9\linewidth]{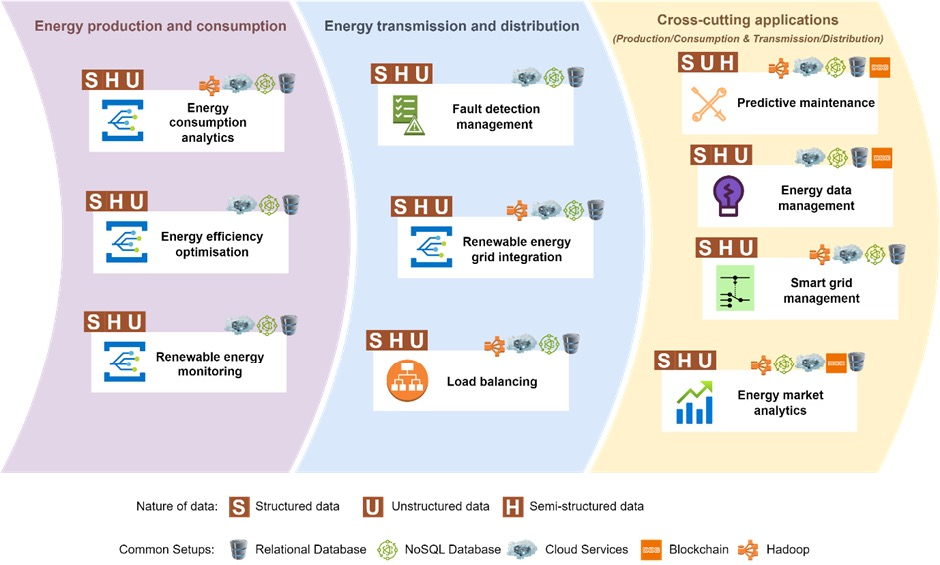}
		\caption{Classification of Data Management Applications in the Energy Sector.}
		\label{fig:classification}
	\end{figure*}

	This section has highlighted an overview of big data management in energy systems and beyond. It is evident that rapid technological advancements and the growing volume, variety, and velocity of energy data highly influence the domain. The state-of-the-art practices have been analysed by narrowing the focus to storage technologies, integration approaches, and their application within the energy sector. Our examination revealed a spectrum of storage technologies and integration methodologies currently in use, each tailored to address specific challenges posed by the volume, variety, and velocity of energy data:
	
	\begin{itemize}
		\item Storage Technologies: Relational databases, NoSQL systems, Hadoop ecosystems, blockchain, and cloud services were discussed in terms of their application and limitations. Whilst relational databases offer robust consistency and integrity, their scalability for big data in energy systems is limited, leading to the adoption of NoSQL for its flexibility with unstructured data. Hadoop has shown promise for large-scale data processing, yet it lacks real-time capabilities. Blockchain introduces decentralisation and security, but struggles with scalability and integration. Cloud services offer a scalable solution for data storage and processing, but they also introduce concerns regarding vendor lock-in and data security. 
		
		\item Data Integration Approaches: The review highlighted traditional ETL and ELT processes, alongside modern methods such as cloud-based data fusion and data spaces. These approaches aim to synthesise data from disparate sources to enhance analytical capabilities, but they face challenges with data volume, real-time processing, and ensuring high-quality data integration. A summary of database management systems currently available in energy systems has been provided in Table \ref{tab:data_management_approaches}. Similarly, an overview of common database system application areas detailing the forms of data involved has been presented in Figure \ref{fig:classification}. 
	\end{itemize}

	{\small 
		\renewcommand{\arraystretch}{0.8} 
		\begin{table*}[t]
			\centering
			\caption{A summary of database management and storage approaches currently available in energy systems.}
			\label{tab:data_management_approaches}
			\begin{tabular}{p{1.5cm}p{1.5cm}p{5.5cm}p{6.5cm}}
				\toprule
				\textbf{Approach} & \textbf{Literature} & \textbf{Applications in energy systems} & \textbf{Key issues/Limitations} \\
				\midrule
				Relational Databases & \cite{RN199, RN446, RN447, RN448} & Energy management information systems; Supervisory control and data acquisition systems (SCADA), grid operation and performance monitoring systems; Modelling energy consumption;Historical data analysis & Scalability constraints; Handling unstructured data; Performance limitation with large-scale data; Integration complexities; Limited flexibility; High maintenance costs; Not ideal for real-time processing \\
				\midrule
				NoSQL Databases & \cite{RN104, RN273, RN274, RN268, RN269, RN267} & Energy consumption data analytics; Smart grids management; Smart meters infrastructure & Lack of standardisation; ACID transactions; Data integration; Query complexity; Limited scalability in centralised settings; Data quality and consistency issues caused by NoSQL’s relaxed consistency model \\
				\midrule
				Hadoop and similar approaches & \cite{RN199, RN206, RN335, RN393} & Smart grids management, including storage and analysis of datasets; Integration of renewable energy sources; Energy consumption monitoring & Limited capabilities with real-time processing; Centralised control; Inefficient with small files; Complex querying; High latency and resource intensity \\
				\midrule
				Blockchain & \cite{RN234, RN255, RN246, RN431, RN423, RN320} & Implementations are scarce but have potential for improving the management of energy systems and efficient data management; Energy tokenisation with smart contracts & Limited scalability as every node processes and stores all transactions; Resource intensive; Limited capabilities with real-time processing; Adopting in the context of energy systems is challenging; Costs and confidentiality concerns; Limited performance \\
				\midrule
				Distributed File Systems, i.e., GFS & \cite{RN173, RN172} & Non-existent & Single point of failure owing to a single master node managing the whole system; Inefficient with small files; Efficient in batch processing hence has limited capabilities with real-time processing; Lacks native support for complex queries and structured data integration; Latency issues when dealing with frequent updates or high-speed data streams \\
				\midrule
				Cloud Approaches & \cite{RN235, RN435, RN436, RN438} & Smart grid management; Energy management systems; Renewable energy integration by managing data from diverse sources; Predictive maintenance; Implementation of demand response strategies & Vendor lock-in and high long-term costs; Data dissipation and fragmentation; Security and privacy concerns; Compliance and regulatory challenges; Challenges integrating with legacy systems; Do not inherently address issues of data quality or redundancy \\
				\midrule
				P2P Data Management & \cite{RN457, bouganim2023highly, gkikopoulos2022decentralised, ameur2024efficient, RN462, RN463} & No case & Maintaining data consistency across replicated nodes in dynamic P2P networks is difficult; Complexity in running distributed queries; Privacy concerns may be complex to address between peers; Overhead for coordinating data replication and synchronisation across nodes; Lack of standardisation; Limited implementation in energy systems \\
				\bottomrule
			\end{tabular}
		\end{table*}
	}

	These findings emphasise the need for innovative frameworks that address data storage, integration, and usability challenges while enhancing scalability and ensuring regulatory compliance. The frameworks should be capable of supporting advanced analytics, secure data sharing, and decentralised architectures to unlock the potential of big data-driven energy systems.
	
	\section{Recommendations}
	\label{sec:recommendation}
	The current state-of-the-art data management approaches present several challenges, yet they still hold significant potential. Uncovering the potential of such approaches in energy systems and beyond is crucial to pave the way for future directions. Generally, the innovation of big data technologies in the energy sector is limited, although it is already taking charge. The highlighted challenges result in fragmented applications and limited data exchange, ultimately leading to the existence of energy data silos. This further limits the ability of stakeholders to share datasets. Here, we highlight four major areas that require improvement in the development of new data management platforms or the advancement of existing platforms.
	
	\subsection{Big Data Storage Solutions}
	\label{subsec:bigdata_storage_solutions}
	Despite the existence of several storage solutions, each at an individual level has advantages and associated challenges. One way to benefit from most of these is to use a hybrid data storage solution that combines the benefits of features of more than one platform. A focus should be put on solutions that can offer better scalability than each in isolation. For instance, a combination of NoSQL or other DBMSs and P2P infrastructure would offer balanced scalability, security, and performance. Several options already exist in the current technology stack. One that has already been implemented in several practices is blockchain technology. Another technology that offers a unique advantage is P2P overlays using DHT for lookup in decentralised storage platforms. P2P architectures are promising for energy systems. Among many other benefits, they hold the potential to enable decentralised energy trading, real-time data sharing among grid operators, decentralised data management, and distributed data analysis in virtual power plants. A distributed storage architecture is necessary to minimise the existing limitations, improve fault tolerance and enhance real-time data access.
	
	The adoption of data lakes would require better governance mechanisms, starting from acquisition and data modelling, to ensure that only data that aid in other services, including Machine Learning studies, is stored. This requires compliance with a common architecture; for instance, having common naming conventions for data objects would enhance data usability and interoperability. This, in turn, would aid in addressing data silos and data swamps, which are already prevalent issues with data lakes and data warehouses, respectively. It is worth noting that priority should be given to platforms that offer compatibility with popular available technologies. Opting for open-source platforms is highly recommended to enhance technological transfer and consequently aid in an effective data-sharing mission.

	\subsection{Data Integration Framework}
	\label{subsec:data_integration_frmwork}
	Recognising that data sharing remains vital among stakeholders, sharing heterogeneous data while maintaining consistency remains a key research concern. Other practices combine multiple clouds, which adds complexity to the integration. In this case, leveraging frameworks like data spaces, energy data reference architecture, and ETL/ELT methods for harmonising heterogeneous datasets is necessary. Additionally, these would support interoperability among systems, compliance with energy sector regulations, and adherence to data management guidelines.
	
	While ETL/ELT would offer integration services, the lookup approaches using DHTs offer an innovative approach to data integration that is hardly explored. With DHT platforms, stakeholders would benefit from their existing DBMSs whilst having the advantage of easily accessing datasets from other systems that share a common architecture. It would also facilitate compliance with existing regulations and data sharing requirements, i.e., the Findable, Accessible, Interoperable, and Reusable (FAIR) principles \cite{RN473}. Developing formalised distributed storage systems that integrate query languages and robust data integration capabilities is essential.
	
	Utilising AI-driven data fusion techniques to enhance the quality and reliability of data integration processes is also a major leap. This would reduce the amount of work that is required to ensure clean data is stored and analysed. These methods can also be integrated on top of storage platforms to enhance new methods of learning, like federated learning. Additionally, in certain applications, the adoption of metadata-driven integration systems, such as the Resource Description Framework (RDF), would facilitate data integration \cite{RN474}. However, they are highly inefficient in many scenarios, e.g. complexities in ontology development, scalability and query performance when dealing with large datasets. One way to realise its benefit is to complement it with other technologies that can partly address the underlying challenges.

	\subsection{Regulatory and best security practices compliance}
	\label{subsec:regulations}
	Regulatory frameworks for data handling have been proposed in \cite{RN264, RN433, RN434, RN210, RN467}. However, given current implementations, the realisation of these frameworks faces several limitations. In this regard, establishing a collaborative framework for data sharing among (common) stakeholders, emphasising trust, data privacy, and compliance with policies like the General Data Protection Regulation (GDPR) is necessary. Some technologies can already facilitate these measures. Blockchain-based technologies would enhance some of the data-sharing rules using smart contracts. However, data quality will be limited. Clouds and Hadoop frameworks can also be adopted since they offer some of these compliances. 
	
\begin{figure*}
		\centering
		\includegraphics[width=.8\linewidth]{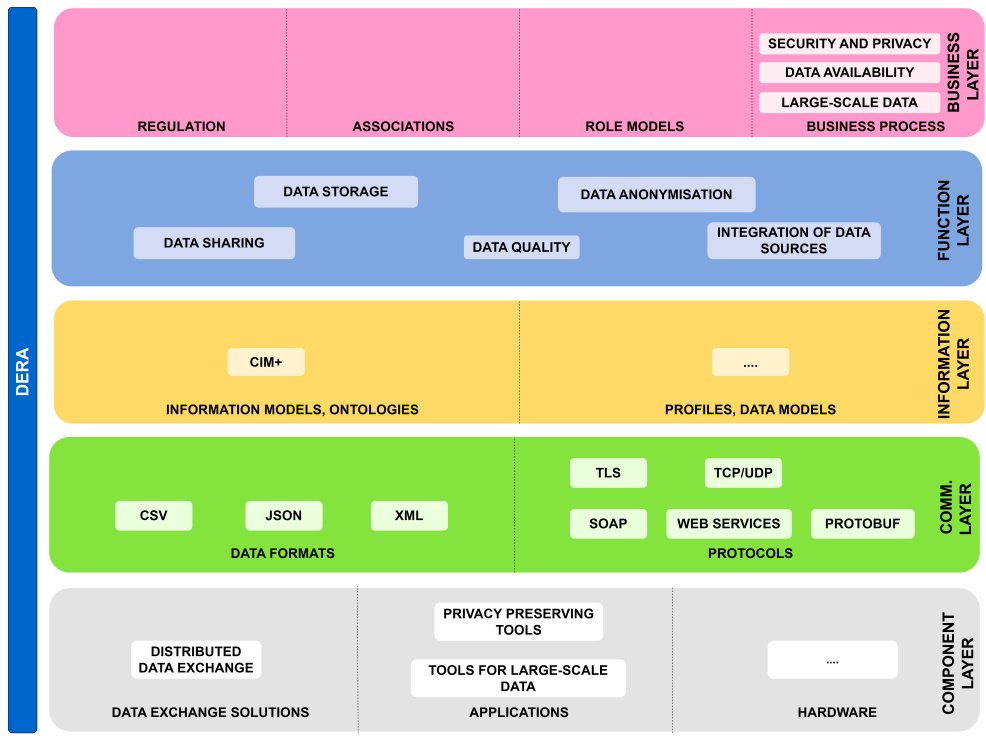}
		\caption{Refined Data Exchange Reference Architecture.}
		\label{fig:red_dera}
	\end{figure*}

	Furthermore, technologies like P2P infrastructure offer a flexible way to implement solutions that comply with the regulatory framework since they offer a high degree of autonomy and participating nodes. Choices would vary according to specific requirements. Our DERA version that would perfectly accommodate the functionalities of P2P systems and DHTs is presented in Figure \ref{fig:red_dera}. An emphasis is put on distributed data exchange approaches and consideration of large-scale data involved while embracing open-source technologies and widely supported platforms, data formats, and protocols. The proposed architecture is also privacy-aware and keen on data security, whilst in store, transmission and integration. 
	
	\subsection{Tailored Solution}
	\label{subsec:solutions}
	There exist several solutions that are not domain-specific, coming from the computer science domain. However, these generic big data solutions are insufficient for addressing the specific needs of the energy sector. There is a pressing need for domain-specific technologies capable of transcending traditional big data capabilities to achieve smart data solutions. While domain-specific solutions are demanded, fostering multi-industry collaboration is equally important. Following open-source implementations and standards would facilitate this. There is also a need for pilot projects that demonstrate the value of innovative big data solutions in renewable energy integration, smart grids, and predictive maintenance for energy systems.
	
	For energy-specific data management solutions, enforcing the common rules for naming entities, data objects, documents, and data files is vital. As a rule of thumb, decentralised data management approaches are compulsory due to the increasing need for scalability. So, the energy sector would have to define its own data naming guidelines by complementing efforts initiated by other previous works and initiatives \cite{RN433, RN434, RN210}.
	
	Security is a critical concern in distributed data storage systems, in which siloed data (commonly found in cloud services and data lakes) poses risks of fragmented and/or incomplete datasets. Employing P2P architectures with secure communication protocols during data transfer offers a promising avenue to enhance data security. Data privacy and confidentiality pose significant barriers, too, particularly when customer data is involved. Energy stakeholders often hesitate to share data due to concerns over business confidentiality, limiting collaborative potential. For sustainable energy systems, consensus-driven approaches are necessary to enable effective data sharing whilst ensuring security and stakeholder trust.

	\section{Conclusions}
	\label{sec:conclusion}
	This work provided a contextual foundation, detailing how big data management in energy systems is driven by the need for analytics, data quality, and integration. It highlighted the challenges of data preprocessing and the role of robust analytics frameworks like Apache, Hadoop, Spark, and Kafka in handling data complexity and scale. The discussion included the management of data uncertainty, which is crucial for applications like load forecasting in energy systems.
	
	With regard to data management, the focus was on various data management practices, from traditional RDBMS to modern NoSQL and distributed databases. The shift from structured to unstructured data management was emphasised, alongside the integration of cloud and blockchain technologies for decentralisation and security. We also covered the evolving landscape of data management in IoT settings, underlining the trend towards decentralised approaches.
	
	The specific needs of the energy sector in the context of data management have also been highlighted. This part evaluated the specific requirements for data management in energy systems, discussing stakeholder efforts toward data exchange and interoperability. Key frameworks like the BRIDGE project's data exchange reference architecture and initiatives like the IDSA's Data Spaces were discussed, highlighting how this aims to facilitate better data utilisation across the sector.
	
	An array of storage solutions was explored, from relational databases to cloud services, focusing on their application in energy systems. The section analysed the limits of each approach — e.g. the scalability issues of relational databases or the consistency challenges with NoSQL — whilst also considering innovative solutions like blockchain and P2P systems for data storage.
	
	With regard to data integration, the discussion revolved around methods and challenges of integrating data from distributed sources. Various integration strategies were outlined, including ETL, ELT, and cloud-based fusion, with specific attention to their effectiveness or shortcomings in the energy sector. The complexity of handling heterogeneous data and the push for real-time integration were key themes, with an emphasis on the need for scalable, secure, and efficient data integration systems.
	
	This work offers a critical examination of data management strategies, setting the stage for future research aimed at addressing the identified gaps and leveraging big data for transformative changes in energy management. The primary contribution lies in its comprehensive analysis of these technologies and methodologies, pinpointing areas where current solutions fall short, particularly in scalability, integration, real-time processing, and regulatory compliance. It advocates the development of domain-specific frameworks that cannot only cope with the specific demands of the energy sector but also foster advanced analytics, secure data sharing, and decentralised system architecture. This work emphasises the need for innovative approaches to unlock the full potential of big data in enhancing energy system operations and strategic decision-making.

	\clearpage
	
	\bibliographystyle{elsarticle-num}
	\bibliography{references}
	
\end{document}